\documentclass[prf,amsmath,amssymb,longbibliography]{revtex4-1}

\usepackage{amssymb}
\usepackage{amsmath}
\usepackage{amsbsy}
\usepackage{bm}
\usepackage{graphicx}
\nocite{apsrev41Control}
\usepackage{changes}

\usepackage{xcolor}

\usepackage[english]{babel}

\usepackage[colorlinks=true, allcolors=blue]{hyperref}

\newcommand{\be}{\begin{equation}}
\newcommand{\ee}{\end{equation}}
\newcommand{\ba}{\begin{eqnarray}}
\newcommand{\ea}{\end{eqnarray}}

\definecolor{red}{rgb}{1,0,0}
\definecolor{yellow}{rgb}{1,1,0}
\definecolor{orange}{rgb}{1,0.5,0}
\definecolor{green}{rgb}{0.,1,0.5}
\definecolor{blue}{rgb}{0,0,1}
\definecolor{white}{rgb}{1,1,1}
\definecolor{purple}{rgb}{0.5,0,0.5}

\begin{document}

\title{Superflows around corners}

\author{Thomas Frisch}
\affiliation{Universit\'e  C\^{o}te d'Azur, CNRS, Institut de Physique de Nice (INPHYNI), 06100 Nice, France}    
\email{thomas.frisch@univ-cotedazur.fr}

\author{Christophe Josserand}
\affiliation{Laboratoire d'hydrodynamique, LadHyX, UMR 7646, CNRS \& Ecole Polytechnique, IPParis, 91120 Palaiseau, France}
\email{christophe.josserand@polytechnique.edu}

\author{Sergio Rica }
\affiliation{Instituto de F\'isica, Facultad de F\'isica, Pontifica Universidad Cat\'olica de Chile, Casilla 306, Santiago,  Chile}
\email{sergio.rica@uc.cl}

%\centerline{v.4 {\today} }

\begin{abstract}
We investigate analytically and numerically  the dynamics of a two-dimensional superflow governed by the Gross–Pitaevski\u\i~ equation passing over finite-size rectangular obstacles: an impenetrable wall and an  impenetrable rectangular well.  Extending classical studies of vortex nucleation around smooth obstacles, we focus on the role of sharp corners 
%and finite aspect ratios obstacle 
 in determining the onset of vortex nucleation. Using a combination of analytical techniques  based on the Schwarz-Christoffel methods for potential flow and on numerical simulations, we show that local velocity amplification near sharp  corners crucially controls the critical flow velocity for vortex nucleation.   
 %The  divergent potential flow at sharp edges is regularized by the  quantum pressure and this lead  to well-defined boundary layer characterized by the healing length.  
 For both wall and well configurations, we identify analytically and theoretically the  critical velocities as a function of the obstacle width and its height or depth, finding an excellent  agreement between the theory and our numerical simulations.
%We find that the critical velocity for vortex nucleation  of a wall increases  with the wall   width  while the converse for the well. Moreover, the asymmetry between the  wall and a well is investigated  analytically and numerically we find that the  critical velocity for  the wall  has a much lower value than the one for the well.
Our results provide a  simple framework for understanding superflow stability past finite-size obstacles with sharp features and are directly relevant to experimentally realizable configurations in atomic Bose–Einstein condensates and related superfluid systems. \end{abstract}
\maketitle
\eject

\nocite{apsrev41Control}

\section{Introduction}

The  existence and the  nucleation of quantized vortices \cite{onsager1949,Feynman1955,Winen1961} is a hallmark of superfluid dynamics and has long served as a probe for understanding the onset of dissipation and drag in systems that are governed by irrotational, inviscid flows.  A simple qualitative description of the physics of vortex nucleation in superfluid helium  can be done by the use of the Gross-Pitaevski\u\i~ (G-P) equation, which is also known as the nonlinear defocusing Schr\"odinger equation \cite{Gross1957,Gross1958,GinzburgPitaevskii1958,Pitaevskii1961,Gross1961,Gross63}. Furthermore, the use of the G-P equation provides quantitative results for atomic Bose-Einstein condensates (BECs) \cite{Ketterle1999}, polaritons in semiconductor microcavities \cite{amo2009superfluidity}, and nonlinear photorefractive optical crystals \cite{Michel2018}. Even though the description of thermal and quantum fluctuations is not present in the G-P equation, this equation captures the essential physics of superflow dynamics thanks to the Madelung transformation that maps it to a compressible fluid augmented by the additional quantum pressure term.
For instance, a fundamental phenomenon captured by the G-P equation is the transition from a steady and laminar superflow to a vortex-laden state when the flow velocity exceeds a geometry-dependent critical threshold. 

The mechanism underlying this transition has been classically studied in references \cite{Frisch1992, Josserand1999} which investigated analytically and numerically superflows passing a circular obstacle and identified vortex nucleation as a nonlinear time-irreversible mechanism that emerges once a local critical velocity at the cylinder apex has been reached. Later on, in-depth numerical investigations of this phenomenon were carried out \cite{HUEPE2000126,Pham2005,Winiecki1999} and led to a more accurate numerical value of the critical velocity for the flow around a cylinder. Furthermore, the effect of the fluid compressibility has been analytically discussed for the flows around a cylindrical obstacle \cite{Rica2001} or a plate-like object \cite{Kokubo2025} by the use of the Janzen-Rayleigh perturbation theory. This analytical method, which relies on the use of a perturbation expansion of a nonlinear phase equation at low Mach number, allows one to improve the predicted value of the critical velocity.
Moreover, and particularly more recently, superflows along very different obstacles have been theoretically studied for geometries such as sinusoidal surfaces \cite{Frisch2024,Josserand1999}, rough surfaces \cite{Stagg2017}, starting flow past an airfoil \cite{Musser2019}, elliptical obstacles \cite{Stagg2014,Stagg2015}, and plate-shaped obstacles \cite{rica1993critical,rica2001vortex,Kokubo2024,Kokubo2025}.
In the G-P equation framework, the obstacles can be modeled as impenetrable barriers using Dirichlet boundary conditions on the obstacle contour as primarily done in \cite{Frisch1992} and \cite{Josserand1999} or as penetrable obstacles using soft potentials \cite{HUEPE2000126,Huynh2024}. In the case of an impenetrable obstacle, due to the quantum pressure term, a boundary layer develops with an intrinsic length scale ($\xi_0$), called  the \emph{healing length}. For penetrable obstacles, an additional physical length which characterizes the potential action and range is added, leading to a non-monotonous behavior of the critical velocity as a function of the barrier height~\cite{Huynh2024}.
Even though the dynamics of superflows along specified boundaries were investigated experimentally and theoretically for classical smooth shapes like disks, much less effort has been made on shapes which contain sharp-shaped angles \cite{rica2001vortex,Kokubo2025}. In fact, it is experimentally more challenging to create obstacles with sharp corners and from a theoretical point of view the presence of sharp corners induces divergence of the potential flow which has to be regularized by the quantum pressure.
In Refs. \cite{rica1993critical,rica2001vortex,Kokubo2024,Kokubo2025} the superflow around a thin plate-shaped obstacle has been studied, and it has been shown that the critical speed decreases when the size of the obstacle increases.

In this Article, we considerably extend the work done in Refs.~\cite{Frisch1992, Josserand1999,Frisch2024,Stagg2017,Musser2019,Stagg2014,Stagg2015,rica1993critical,rica2001vortex,Kokubo2024,Kokubo2025} to new geometries: a rectangular barrier (wall) and a rectangular well embedded in a homogeneous two-dimensional superfluid flow.  
In the work of Ref. \cite{Kokubo2025} the obstacle was a thin plate with just one characteristic length while in our case, we consider finite-size objects characterized by two lengths, their finite width and their finite height for a wall or a well.
The motivation for this extension is twofold. First, such geometries offer an opportunity to explore how features like sharp corners influence the superfluid local velocity and thereby modify the conditions for vortex nucleation. In particular, it will
provide a well defined geometry to investigate how the singular behavior of the velocity when considering incompressible potential flow, interacts with the boundary layer induced by the quantum pressure regularization near the wall. 
Secondly, realistic physical and experimental configurations often involve non-cylindrical obstacles such as steps, channels, or potential wells which could be generated by tailored optical or magnetic fields in BEC systems.
We thus investigate objects with sharp corners and of finite size in order to capture the effect of the geometry of the obstacle and the effect of the value of the angle of their sharp corners.

Our Article is organized as follows.
We first present in Section \ref{Sec:Model} our model which is based on the G-P equation and we precise the obstacle geometry. We then perform in Section \ref{Secc:Numerics} two-dimensional numerical simulations showing how vortex nucleation happens around these obstacles. In Section \ref{Sec:CriticalVelocity} we first recall the theoretical framework for vortex nucleation using the G-P equation, we then show how to derive analytically the critical value for vortex nucleation for our specific geometry by means of a complex mapping based on the Schwarz-Christoffel method. Moreover, we perform an asymptotic development that unifies the expected results for the critical speed around a corner with the known results for the critical velocity passing a thin plate-shaped obstacle \cite{rica1993critical,rica2001vortex,Kokubo2024,Kokubo2025}.
In Section \ref{Sec:Results} our results are confronted with our numerical simulations for different system sizes and an excellent agreement is found. In accordance with our theory, our numerical simulations show that the critical velocity for vortex nucleation of a wall increases with the wall width while the converse holds for the well.
In Section \ref{Sec:Conclusion} we conclude this Article and we provide some future perspectives.

\section{ Basic Model and geometry}\label{Sec:Model}

\subsection{The Gross-Pitaevski\u\i~ equation.}

To model superflows, we use the Gross-Pitaevski\u\i~ equation (GP)~\cite{Gross1957,Gross1958,Gross1961,Gross63,GinzburgPitaevskii1958,Pitaevskii1961} that reads~:
\be
i \hbar\frac {\partial \psi} {\partial t} =  - \frac{\hbar^2}{2m} \nabla^2  \psi + g  |\psi|^2\psi ,
\label{eq:NLSorg}
\ee
where, $m$ is the  mass of particles, $2\pi\hbar$ is the Planck constant, and, $g=\frac{4\pi \hbar^2}{m}\times  a_0$ is the interaction characterized by the scattering length, $a_0>0$, in the limit of low energy collisions.  Finally, $\psi$ is a complex scalar field depending on time $t$ and on two-dimensional  space ${\bm x}=(x,y)$, whose modulus represent the number density of particles, that is $|\psi|^2$ has units of the inverse of a volume, and its phase will be interpreted as a velocity potential (see below).

 A remarkable characteristic of the Gross-Pitaevski\u\i~ equation, Eq.~(\ref{eq:NLSorg}),  is the time reversibility  (if $ \psi ({\bm x}, t) $ is solution then $ \psi ^ * ( {\bm x}, -t) $ is also a solution). This is due to the Hamiltonian structure of the equation: defining the time conserved energy functional $H$ by
\begin{equation}  H = \int \left (  \frac {\hbar^2} {2m}  |  {\bm \nabla} \psi | ^ 2 + \frac {g} {2} | \psi | ^ {4} \right) d {\bm x}  \quad , \label{eq:Hamiltonian}
 \end{equation}  
we can express  Eq.~(\ref{eq:NLSorg}) as: 
 $i \hbar \frac {\partial \psi} {\partial t} = \frac {\delta H} {\delta \psi ^ *} .$ Among other relevant features, the Gross-Pitaevski\u\i~ model admits a hydrodynamical interpretation thanks to the Madelung transformation. By setting $\psi = \rho^{1/2} e^{i\varphi}$, we can
transform Eq.~(\ref{eq:NLSorg}) into a coupled set of hydrodynamic-like equations:
\begin{eqnarray}
\partial_t \rho & =  &- {\bm\nabla}\cdot\left( \rho \frac{\hbar}{m} {\bm\nabla}\varphi\right) \, ,  \label{Eq:cont} \\
\hbar\partial_t \varphi & = & \frac{\hbar^2}{2m\sqrt\rho} \Delta \sqrt{\rho}  
- \frac{\hbar^2}{2m}({\bm\nabla} \varphi)^2 - g \rho({\bm x}) \, .
\label{Eq:Bernoulli}
\end{eqnarray}
 Here, $\rho = |\psi|^2$ is a density field and, 
 \begin{equation}
 {\bm v} = \frac{\hbar}{m} {\bm \nabla}\varphi,
\label{Eq:SuperfluidVelocity}
\end{equation}
is  the superfluid velocity field. Eq.~(\ref{Eq:cont}) corresponds to a continuity equation
for the density $\rho$, and   Eq.~(\ref{Eq:Bernoulli}) is a 
Bernoulli-like relation for the velocity potential $\frac{\hbar}{m}\varphi$.
The first  term on the right of the equation (\ref{Eq:Bernoulli}) is called the ``quantum pressure" because it vanishes in the classical limit ($\hbar\to 0$).
Thus, without this quantum pressure term,  Eqs.(\ref{Eq:cont}) and (\ref{Eq:Bernoulli}) represent a classical compressible fluid with the following  equation of state:
$$ p = \frac{g}{2m}  \rho^2,$$
so that the Gross-Pitaevski\u\i~ equation naturally admits sound waves in the long-wave limit. More specifically, perturbations of density $\rho({\bm x},t)$ and phase $\varphi({\bm x},t)$ around a uniform background at rest, 
 ($\rho=\rho_0 $ constant, $ \nabla \varphi_0 =0$),
 propagate as dispersive waves. The dispersion relation obeys the well known Bogolyubov spectrum \cite{Bogolyubov1947} which in the long-wave limit becomes an acoustic mode with the speed of sound
  $$  c =\sqrt{\left. {\partial p}/{\partial \rho} \right|_{\rho_0}}=\sqrt{\frac{g}{m} \rho_0}= \frac{\hbar}{m} \sqrt{4\pi a_0 \rho_0}.$$
  
In this configuration, a typical scale, the so-called healing length, emerges as a balance between the kinetic and interaction energy in GP, that is
$$ \xi_0 = \frac {\hbar} {mg \sqrt{\rho_0}} = \frac{1}{4\pi a_0 \rho_0}.$$
This length is therefore related to the quantum pressure term and provide the scale of the boundary layer separating an impenetrable body with the far field density $\rho_0$.

In the following, we will use the  dimensionless
Gross-Pitaevski\u\i~ equation, often called also the Non-Linear Schr\"odinger equation (NLS). Defining:
\begin{eqnarray*}
{\bm x}' = {\bm x}/\xi_0, \quad 
t'= c t/\xi_0 , \quad
\psi'= \frac{1}{\sqrt{\rho_0}} \psi,
\end{eqnarray*}
we obtain (omitting the primes for simplicity):
\be
i \frac {\partial \psi} {\partial t} =  - \frac{1}{2} \nabla^2  \psi +  |\psi|^2\psi  \,,
\label{eq:NLS0}
\ee

It is worth noting that in this dimensionless equation, the density of the fluid at infinity is now simply $\rho_0=1$.
Finally, to model an impenetrable obstacle, as in \cite{Frisch1992,rica1993critical,Winiecki1999,rica2001vortex}, we set the wave function $\psi(x,y)=0$ at the boundaries. Additionally, to model a flow with constant velocity $v_0 \hat{\bm x}$ and constant density $\rho_0$ at infinity, we operate a Galilean boost of the superflow, keeping the boundary of the obstacle fixed. This is done by making the transformation $ \psi(x, y,t) \to  \psi(x, y,t) e^{i (v_0 x- \frac{1}{2} v_0^2 t)}$. Therefore, the flow around an obstacle will be studied in the framework of  the following NLS equation~:

\be
i\partial_t \psi= -i  v_0\frac{ \partial \psi}{\partial x}  - \frac{1}{2} \nabla^2 \psi +  |\psi|^2\psi \, .  \label{eq:NLS}
\ee
Within this framework, we can now investigate numerically and theoretically flow around obstacle by imposing $\rho_0=1$ at infinity and $\psi=0$ on the obstacle boundaries.

\subsection{Obstacle shapes}\label{subSecc:Obstacles}
\begin{figure}[h!].
\centering
\includegraphics[width=0.8\columnwidth]{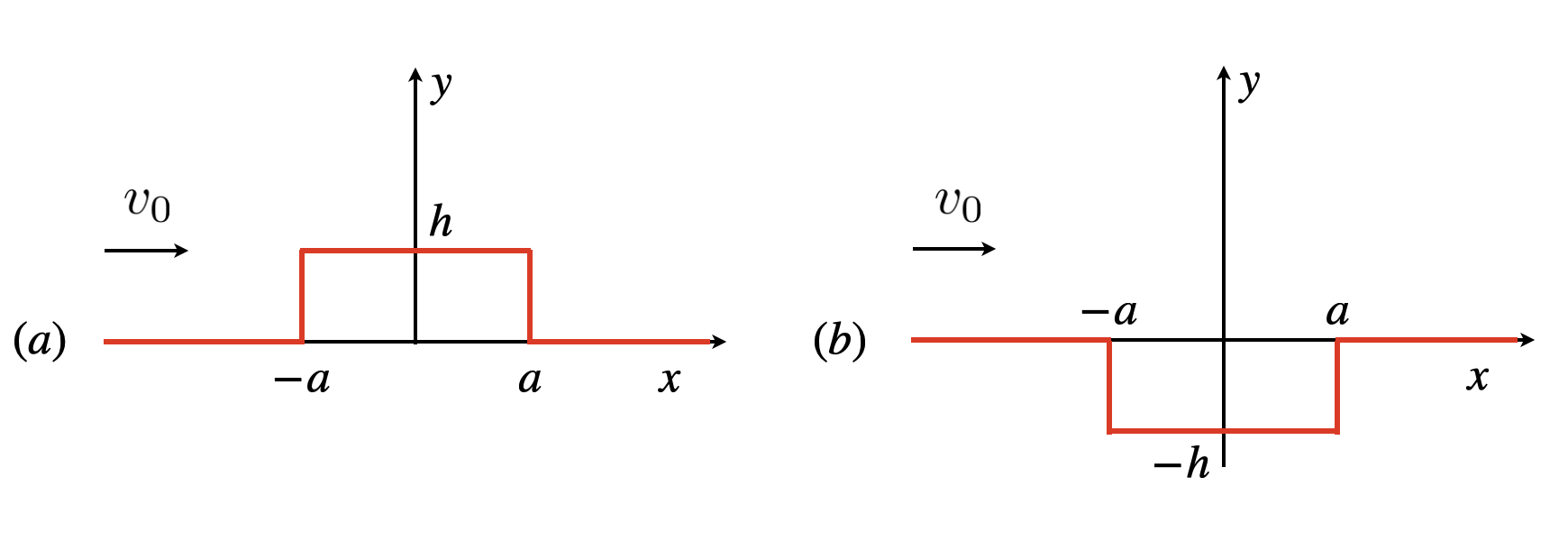} 
\caption{\label{fig:Fig1} A wall of height $h$ and a well of depth $h$. Both domains have a width $2a$. }
\end{figure}

In the present study, we thus focus on the flow around two objects of finite size with sharp corners, a barrier of height  $h$ and base  $2 a$ and, a well of depth $h$ and base  $2 a$ (as shown in Fig. \ref{fig:Fig1}).
These geometries have the advantage of simplicity and allow an analytical derivation of the irrotational potential flow, using the Schwarz-Christoffel transformation which provides their conformal complex mapping \cite{milne1996theoretical}.  
As we will show, the presence of such sharp obstacle in a uniform flow leads for potential flows to singular behavior
Indeed, as known from the classical potential theory of fluid dynamics around
sharp objects \cite{milne1996theoretical}, the local velocity diverges as $v\sim r^{-1/3}$, $r$  being the distance to the corner, for the sharp corner with an opening of  $3\pi/2$ ({\it e.g.} the corners at $x=\pm a$ and $y=h$ in Fig. \ref{fig:Fig1}-a, and those at $x=\pm a$ and $y=0$ in Fig. \ref{fig:Fig1}-b). 
On the other hand, for the case of the acute corner with an opening of $\pi/2$ ({\it e.g.} the corners at $x=\pm a$ and $y=0$ in Fig. \ref{fig:Fig1}-a, or at $x=\pm a$ and $y=-h$ in Fig. \ref{fig:Fig1}-a), the velocity flow corresponds to a stagnation point with a  velocity vanishing linearly, $v\sim r$. 
However, in the Gross-Pitaevski\u\i~ framework, which models a compressible fluid, this singularity is regularized by the quantum pressure due to the boundary conditions at the wall, on a distance of the order of the healing length $\xi_0$. 
Finally, we thus expect vortex nucleation near the sharp corners only and for velocities $v_0$ bigger than a critical velocity $v_0^c$ that is determined by the interplay between the potential singular flows and the boundary layer. The goal of our paper is thus to
compute numerically and model theoretically how this critical velocity varies as $a$ and $h$ do.
As we shall see, the main difference between the barrier and the well is that the critical velocity increases when $a$ increases in the barrier case, while it decreases for the well.

\section{Direct numerical simulations of a superflow in the framework of the Gross-Pitaevski\u\i~ equation}\label{Secc:Numerics}

 We thus perform numerical simulations of  Eq.~(\ref{eq:NLS}) using a finite difference method, which is well adapted to describe sharp corners. To simulate the flow, we impose a constant density far from the object $|\psi_0|^2=1$ (which implies that both  the healing length $\xi_0$, and the speed of sound $c$, remain fixed at the unit value), and, a uniform velocity $v_0 \hat{\bm x}$ at infinity.
In fact, since numerical simulation are done over a finite domain of size $L_x \times L_y$, we impose periodic boundary conditions in the $x$ direction, while we use  Dirichlet boundary conditions, $\psi(x,y)=0$ on the bottom wall (the obstacle) and Neuman boundary conditions at the top $\left.
\left( \frac {\partial \psi}{\partial y}\right|_{y=L_y} =0\right)$. Since  we use periodic boundary conditions, a wall of width $2a$ becomes a well of width $L_x -2a $ as soon as $a > L_x/4$.  This property provides a convenient method to study both cases in a single run. However, it has the numerical limitations of finite-size periodic system making more subtle the comparison between the numerics and the theory based  on compressible fluid  flows (Secc. \ref{Sec:CriticalVelocity}). A possible improvement for the comparison between the theory and the numerical would be to use periodic Schwarz-Christoffel mapping for the theory, as developed for example in Ref. \cite{Baddoo2019} but  this is beyond the scope of our article.

\begin{figure}[h]
\begin{center}
\centerline{(a) \includegraphics[width=5cm]{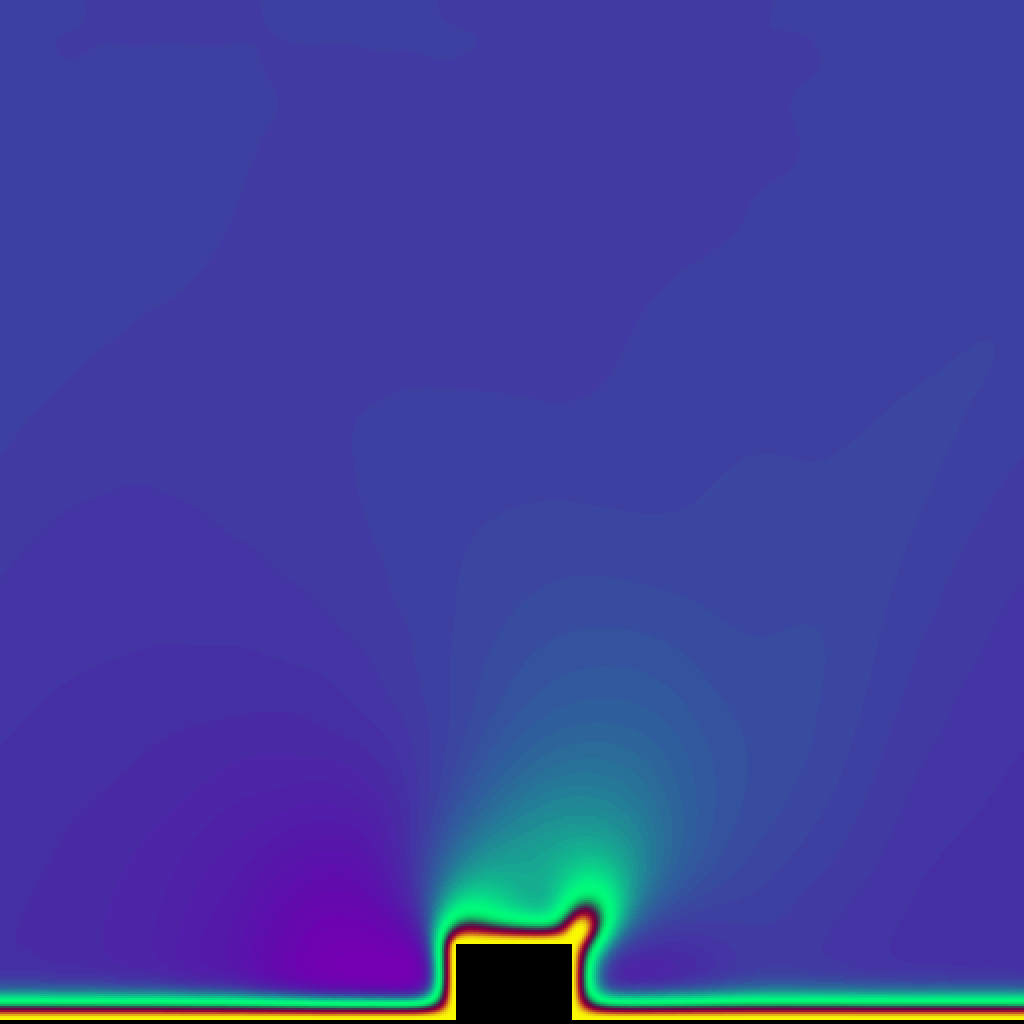} \,  (b) \includegraphics[width=5cm]{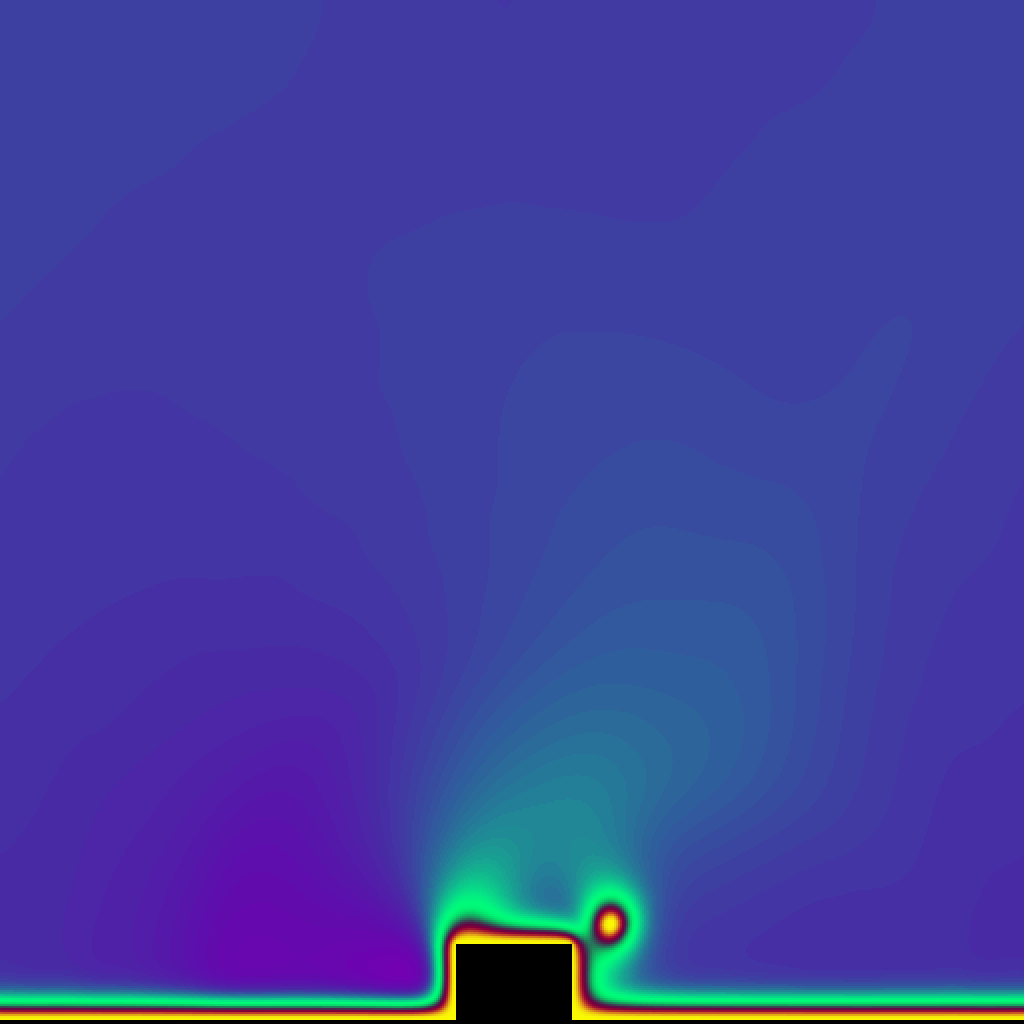}\, (c) \includegraphics[width=5cm]{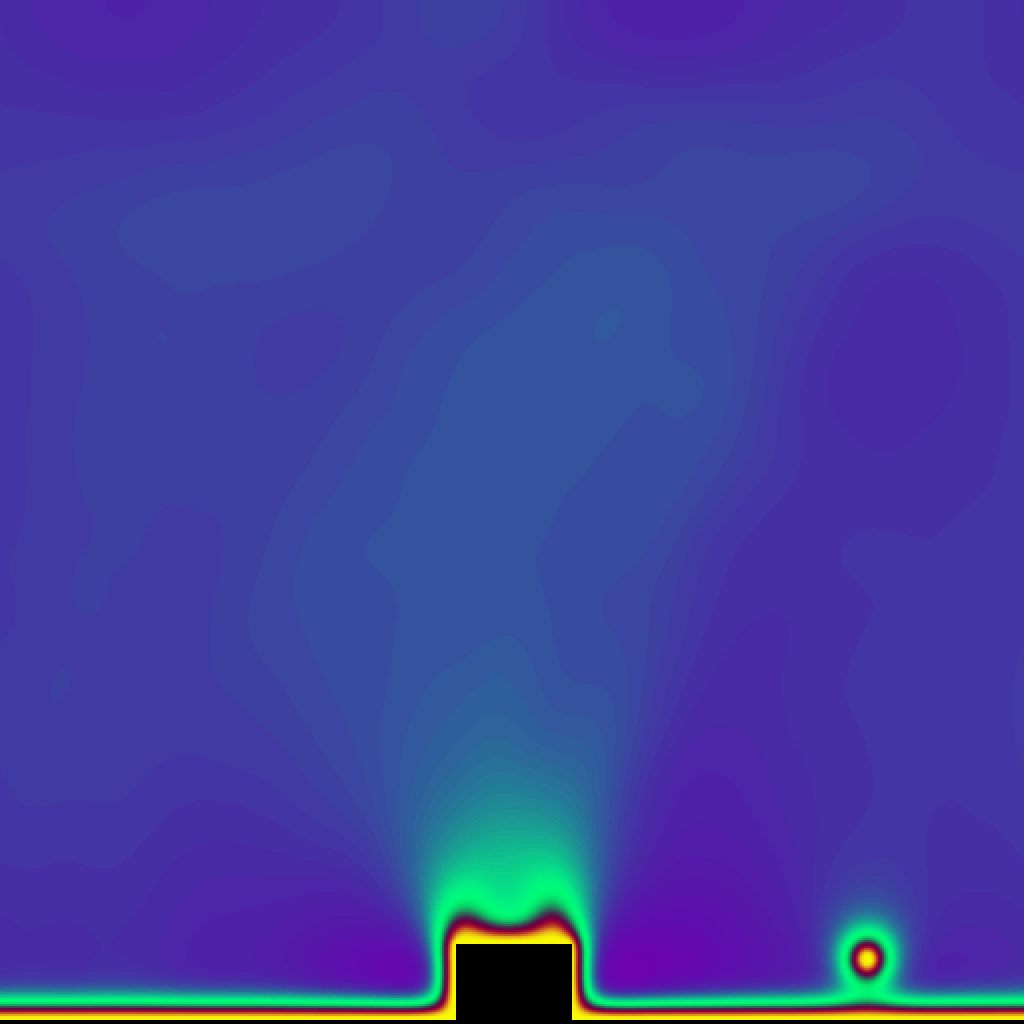}}
\centerline{(d) \includegraphics[width=5cm]{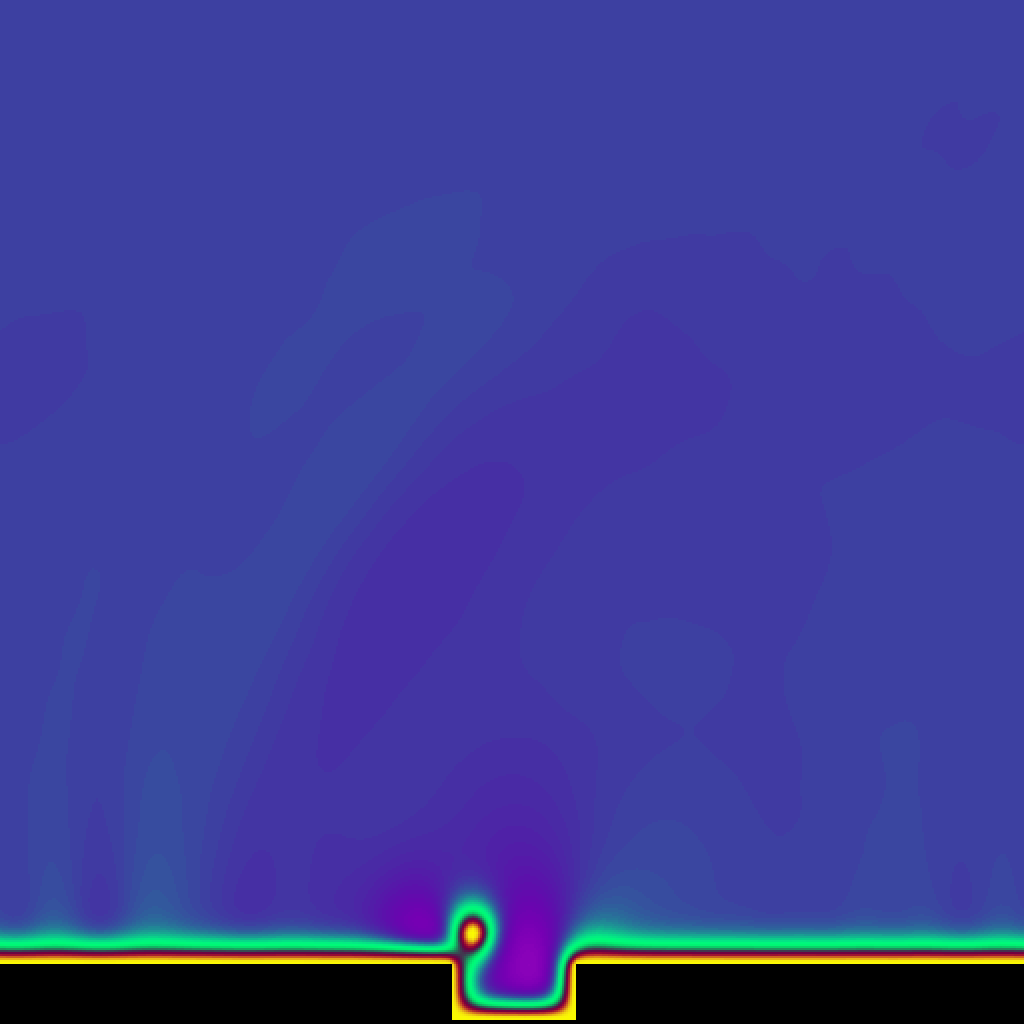} \,  (e) \includegraphics[width=5cm]{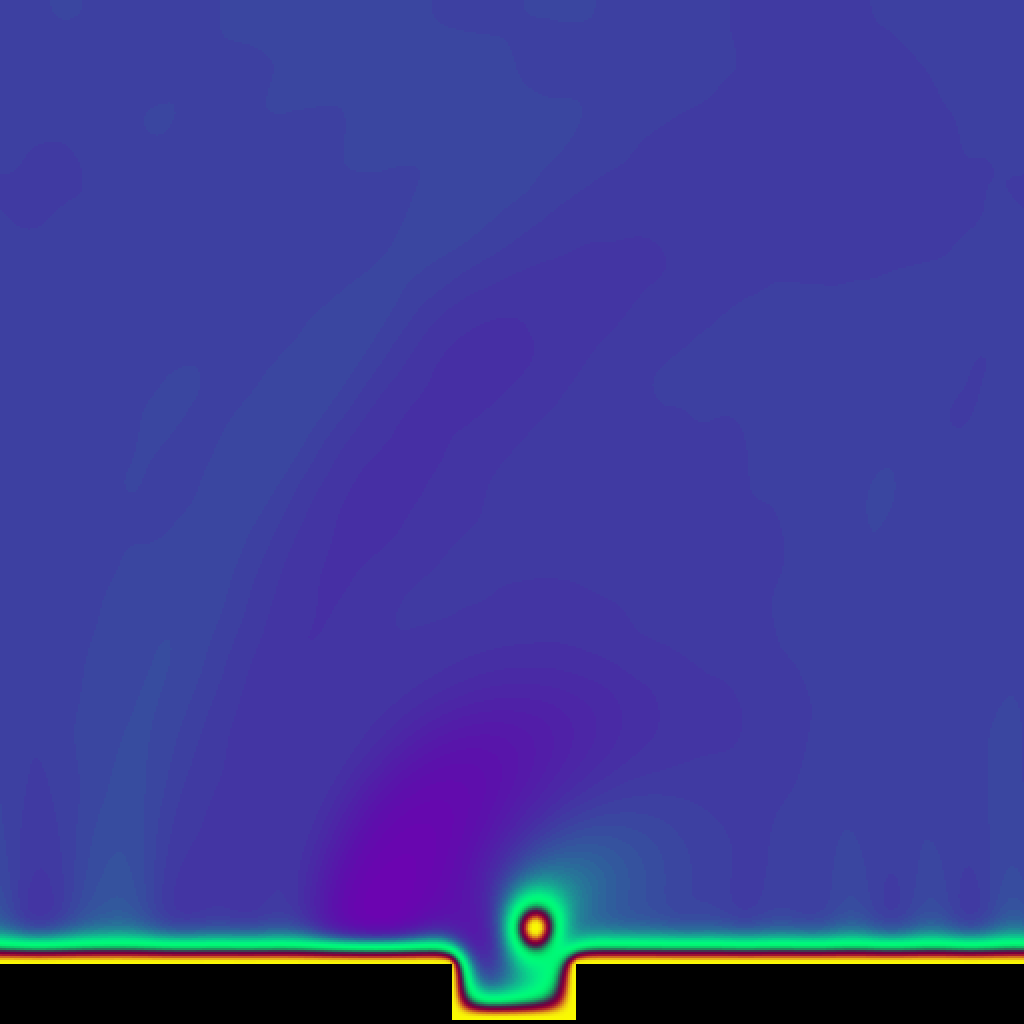}\, (f) \includegraphics[width=5cm]{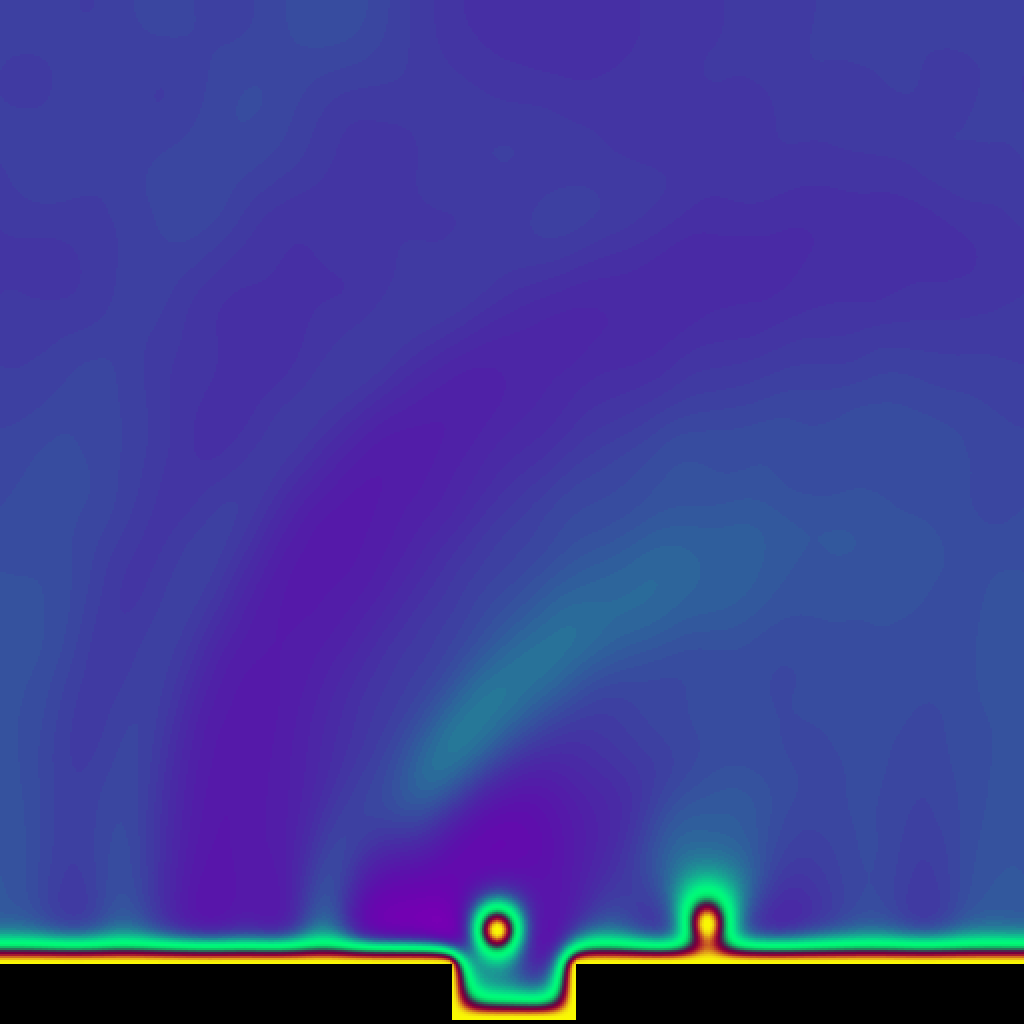}}
\caption{ \label{Fig:NumericsWall} 
 Numerical simulation of the superflow passing around a wall (top) and a well (bottom) and displaying vortex nucleation. The upper and lower snapshots display the density field, $|\psi|^2$, for three increasing times. In both cases the numerical simulations are for $|\psi_0|^2=1$, $\xi_0=1$, the space and time discretization are $ dx= 0.25 $ and $dt =0.0025$, and the system size is $64\times 64 $. The obstacle is in black, the blue corresponds to $|\psi|^2\approx 1$, and the yellow matches with $|\psi|^2\gtrsim 0.$
Top: The parameters are the Mach number $  M=0.5,$ the obstacle width $2a= 7.5,$ and its height  $h = 5$. Bottom: The parameters are $  M=0.725 ,\, 2a= 7.5, \, h = 3.75 $. Notice that in both cases, vortex are nucleated near the right open corners in the sense of the flow, as it can be seen in the snapshots (b) and (d). Subsequently, the vortices are advected in the direction of the flow, as it is shown in the snapshots (c) and (f).
}
\end{center}
\end{figure}

To achieve the prescribed velocity, we first start the simulation with $v_0=0$ and relax the solution to a minimum of  the energy of Eq.~\eqref{eq:Hamiltonian} with the constraint that $|\psi|^2=1$ far from the obstacle. This is done by simulating:
\be
\partial_\tau \psi=  \psi + \frac{1}{2} \nabla^2 \psi -  |\psi|^2\psi \, ,
 \label{eq:StationaryGP}
\ee
 for a transient ``time'' $\tau=10$ up to $50$ units which ensures a good initial condition close to the ground state. Next, the dynamics is shifted to the conservative scheme given by Eq.~\eqref{eq:NLS}, and the flow velocity is quasi-statically increased, for $200$ time units, up to a desired velocity $v_0$.
As we increase the velocity $v_0$, we observe that the stationary regime disappears above a critical velocity $v_0^c$. For $v_0 >v_0^c$ quantized vortices are released at the edge of the $3\pi/2$ corner, for both the wall and the well geometries (See Fig.  \ref{Fig:NumericsWall}-(b) \& (d)). Moreover, as we increase the velocity toward larger values above the critical velocity, we observed  that   the number of  vortices  increases and this may lead to a complex quasi turbulent  2 dimensional  states which is beyond the scope  of the paper and  requires the tools of  a statistical approach  and an ensemble average.

Amazingly, since the solution for  the potential flow is symmetric  with respect  to  the horizontal coordinate ($x\to -x$), the velocity field is exactly the same near the corner at  $x=+ a$ and the corner at  $x=- a$. We should therefore expect the nucleation of vortices to happen in both side of the obstacles, which is not observed in the numerics! We suspect in fact that the spatial reversibility of the flow ($x\to -x$) is broken by the quasi-statically increase of the velocity . As shown in Fig \ref{Fig:NumericsWall}(a)-(b)-(d), the vortices are  nucleated on the downstream corner edge of the wall  ($x= i a, \, y=h$) and  the upstream corner of the well  ($x= -i a,\,  y=-h$).
Finally, our numerical simulations reveal also that the critical velocity decreases as the height of the barrier increases for a fixed obstacle width ($2a$). This happens for both cases, the wall and the well. On the other hand, the critical velocity  increases as $a$ increases, for a fixed value of the wall height, $h$, while it decreases for the well.
In what follows, we will thus investigate quantitatively how $v_0^c$, corresponding to a critical Mach number $M_c=v_0^{c}/c$ (since $c=1$ here) varies with $a$ and $h$. 
The numerical results are presented in the forthcoming Section \ref{Sec:Results}, while the next section \ref{Sec:CriticalVelocity} develops an analytical prediction for $M_c=v_0^{c}/c$.

\section{General Theory for the Critical Velocity for vortex nucleation}\label{Sec:CriticalVelocity}
\subsection{The transonic transition}

It was shown in Ref. \cite{Frisch1992} that the phenomenon of vortex nucleation in the Gross-Pitaevski\u\i~ equation is linked to the problem of a transonic transition. It means that for some critical up-stream velocity, $v_0^{c}$, the local speed on or near the obstacle surface becomes supersonic locally.
Because the Gross-Pitaevski\u\i~ equation does not have any intrinsic dissipation, as it is the case for shock waves in classical compressible fluids, no such dissipation can regularize  sharp gradients induced by the supersonic transition. 
 Instead, the sharp gradients are regularized by  dispersive effects due to the quantum pressure presence. This leads to the creation of dark solitons forming a dispersive shock. These dark solitons, in which the modulus vanishes are the precursors of the vortices. In the present case,  the presence of sharp obstacles in a uniform flow increases necessarily the local speed at the corners  of the obstacle, implying a drastic diminution of the  critical flow velocity as shown in  \cite{PomeauRica1993,rica1993critical,Kokubo2024,Kokubo2025}. 
 
We  now describe carefully the mathematical treatment for the determination of the critical velocity.
In Refs. \cite{Frisch1992,Rica2001} the problem of the critical velocity is reduced in the limit of zero quantum pressure to the compressible flow described by the stationary problem and we will put our feet in the same prints.
This problem matches perfectly with the stationary motion of compressible fluids \cite{Landau1987Fluid}.  This analogy comes directly from the stationary solution of Eq. (\ref{Eq:cont}), together with the (trivially) time-dependent solution of Eq.~(\ref{Eq:Bernoulli}) that satisfies correctly the boundary conditions: $\rho({\bm x}) \to \rho_0=1$ and  ${\bm\nabla }\varphi \to v_0\,\hat{\bm x}$ as $|{\bm x}| \to\infty$.  Thus, the phase obeys the relation
\begin{eqnarray} \varphi = - \left( \rho_0 + \frac{1}{2} v_0^2\right)t + \phi({\bm x}),\label{eq:DefPhi}
\end{eqnarray}  
where $ \phi({\bm x})$ does not depend explicitly on time, with $ \phi({\bm x})\to v_0 x$ as $|{\bm x}| \to  \infty$.
Neglecting the quantum pressure term $ \frac{1}{2\sqrt\rho} \Delta \sqrt{\rho} $ in  Eq.~(\ref{Eq:Bernoulli}) provides a Bernoulli-like relation for $\rho({\bm x})$ in terms of  $v = \sqrt{ \left({\bm \nabla } \phi \right)^2 }$:
\begin{eqnarray}
\rho(v) =  c^2+ \frac{1}{2} v_0^2 -  \frac{1}{2} v^2. \label{Eq:BernoulliLike}
\end{eqnarray}
Therefore, from Eq. (\ref{Eq:cont}) $\phi$ is governed  by the following nonlinear partial differential equation: 
\begin{eqnarray}
{\bm \nabla }\cdot \left(  \left(  c^2+ \frac{1}{2} \left(v_0^2 -   \left({\bm \nabla } \phi \right)^2 \right)\right) {\bm \nabla } \phi \right) &=& 0  \quad {\in \Omega}, \label{eq:Continuity} \\
{\bm \nabla } \phi \cdot \hat {\bm n}&=&0 \quad {\in \partial\Omega} , \label{eq:Boundary1} \\
{\bm \nabla } \phi &=&v_0\, \hat {\bm x} \quad {\rm when } \quad {(x,y) \to \infty}  .\label{eq:Boundary2}
\end{eqnarray}
where $\Omega $ represents the domain, and 
$
\hat{\bm n}$
 is the unit vector perpendicular to this boundary.

As a general result, for $\phi$ solution of Eq.~(\ref{eq:Continuity}) with the boundary conditions \eqref{eq:Boundary1} and \eqref{eq:Boundary2}, the maximum velocity $\left|{\bm \nabla } \phi \right|_{\rm max}$ is reached on the boundary of the obstacle and, in general, $\left|{\bm \nabla } \phi \right|_{\rm max}> v_0.$ 
Moreover, as $v_0$ increases, $\left|{\bm \nabla } \phi \right|_{\rm max}$ increases until a critical value,
$ v^{(c)}_0
$,
 for which the elliptic  Eq.~(\ref{eq:Continuity}) becomes hyperbolic as soon as the local speed $\left|{\bm \nabla } \phi \right|_{max}$ reaches $v_*$, a local critical velocity, as explained below, following~\cite{Landau1987Fluid,Frisch1992}. 
Firstly, coupling the Bernoulli-like relation (\ref{Eq:BernoulliLike}), which links the density to the fluid velocity with the continuity  Eq.~(\ref{eq:Continuity}) we obtain:
\begin{eqnarray}
\rho\left( v  \right)  { \nabla } ^2\phi  + \frac{\partial \rho(v)}{\partial v}  \frac{\partial_i\phi   \partial_k\phi }{v}   \partial_{ik}\phi &=& 0. \label{eq:Continuity2}
\end{eqnarray}
Here $v= \left|{\bm \nabla } \phi \right| $, and we have used  $\partial_i \rho(v)\equiv \partial_i \rho\left( \left|{\bm \nabla } \phi \right|  \right)  = \frac{\partial \rho(v)}{\partial v} \partial_i v= \frac{\partial \rho(v)}{\partial v} \frac{v_k}{v} \partial_iv_k$, with $v_k =  \partial_k\phi $ (we use the Einstein notation where repeated index stand for the sum). 

Writing  Eq.~(\ref{eq:Continuity2}) in a diagonal form, that is by setting a local set of variables that suppresses second-order cross derivatives, we obtain that
 Eq.~(\ref{eq:Continuity2}) is {\it a priori} an elliptic differential equation, that can however become hyperbolic when locally  $d (v\rho(v))/dv= ( \rho\left( v  \right) + v\rho'\left( v  \right)) $ vanishes and changes sign for a critical  value of the velocity $v= v_*$ \footnote{Although $\rho(v) $ may also changes of sign, in our case of interest $( \rho\left( v  \right) + v\rho'\left( v  \right)) $ changes sign  first. }.
Therefore, the local critical velocity, $v_*$  which is a function of  the upstream velocity $v_0$, satisfies 
$$
 \rho\left( v_*  \right) + v_*\rho'\left( v_*  \right)= c^2+ \frac{1}{2} v_0^2  - \frac{3}{2} v_*^2 = 0 ,
$$
that is \cite{Frisch1992}, 
\begin{equation}
 v_*^2 = \frac{2}{3} c^2+ \frac{1}{3} v_0^2.  
\label{eq:criticalvelocitysquare}
\end{equation}
Here, the local critical  $v_*$ should not be confused with the upstream critical velocity $v_0^{c}$, which we have defined earlier after Eq.~(\ref{eq:Boundary2}) and $v_0^{c}$ is the upstream critical velocity at which the local fluid velocity  $v_*(v_0)$  which is a function of $v_0$ becomes transonic, thus $v_0^c$ obeys the  relation Eq,~\eqref{eq:criticalvelocitysquare} with $v_0^c$ instead of $v_0$.

In the case of the flow around a cylinder, we remind the reader that by means of the classical incompressible potential theory $v_*(v_0)= 2 v_0$ at the cylinder  poles \cite{milne1996theoretical}, so 
that  Eq. (\ref{eq:criticalvelocitysquare}) gives $v_0^c= \sqrt{\frac{4}{11}} c$ \cite{Frisch1992}.

\subsection{Incompressible flow  passing around a wall domain}
 In Ref.~\cite{Frisch1992}, the critical velocity has thus been estimated  through the compressible critical velocity conditions (\ref{eq:criticalvelocitysquare}) using an inviscid incompressible approximation of the flow. 
 Proceeding similarly, in this section we compute such zero-th order inviscid potential solution by using a conformal mapping. Because the domains are of polygonal shape, we can use a Schwarz-Christoffel mapping \cite{milne1996theoretical,Carrier2005} to deduce 
 analytical expression of the flow.

\begin{figure}[!htp]
\begin{center}
\includegraphics[width=12cm ]{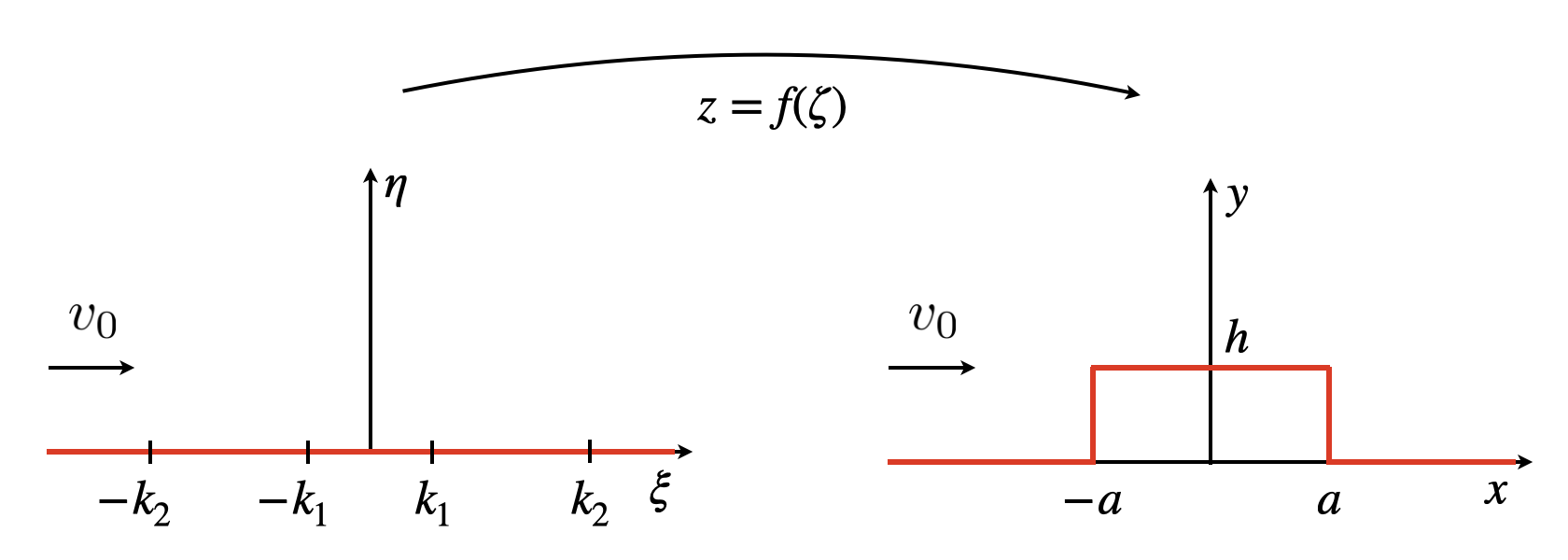}
\caption{ \noindent Conformal map $z=f(\zeta)$ for transforming the plane $\zeta$-plane into a wall in the $z$-plane. Here $f(0)= ih$, $f(\pm k_1)= \pm a+ ih$, and $f(\pm k_2)= \pm a$. }
\label{Fig:Wall}
\end{center}
 \end{figure}

In order to solve the flow along a wall or a well (see Fig. \ref{fig:Fig1}) we shall use the  hydrodynamic potential theory which uses a 
complex mapping method which  maps the complex physical
plane $z=x+i y$ into the $\zeta= \xi+i \eta$  plane \cite{milne1996theoretical,Carrier2005}. In the $z$  plane, the geometry is complex (a rectangular wall)  so that the Laplace equation is hard to solve {\it a priori}
In  the $\zeta$-plane, the  geometry is simple  and Laplace equation can  be easily solved analytically.
By contrast, in the  $\zeta$-plane, we consider a simple uniform flow parallel  to a flat boundary, so that the complex velocity potential writes:
\begin{equation}
\chi(\zeta ) =\Phi(\zeta)+i \Psi (\zeta) = v_0 \zeta. \nonumber
\end{equation}
Here $\Phi$ is the velocity potential and $\Psi$  is the stream  function.
Here $\Phi= {\rm Re}(\chi(\zeta ))$ and  $\Psi= {\rm Im}(\chi(\zeta ))$
thus $\Phi= v_0 \xi $ and $\Psi= v_0 \eta$.
In the  $\zeta$ plane the hydrodynamic flow, $V_{\eta}=\frac{\partial  \Phi}{ \partial\eta }=0$, is just constant and parallel to the  $\xi$ axis $$V_{\xi}=\frac{\partial  \Phi}{ \partial\xi}=v_0 \quad  {\rm and} \quad   V_{\eta}=\frac{\partial  \Phi}{ \partial \eta}=0,$$
and equivalently $ V= V_{\xi} - i V_{\eta}= \frac{d \chi(\zeta)}{d\zeta}=v_0$.
We now introduce the conformal  mapping $z=f(\zeta)$ as shown  in Fig. \ref{Fig:Wall} which maps a straight line to a rectangular wall.
We recall that the  function $\zeta^{1/2}$  draws a $\pi/2$ angle when mapping the $\xi$  axis from $-\infty$ to $\infty$, thus a rectangular barrier can be easily constructed by four  $\pi/2$  angle corners with two of them of opposite sign.

This mapping  can be defined in a differential form  \cite{milne1996theoretical,Carrier2005} as follows:
\begin{eqnarray}  \frac{dz}{d\zeta} =  \sqrt{\frac{\zeta^2-k_1^2}{\zeta^2-k_2^2}}=f'(\zeta) . \label{eq:SchChr} \end{eqnarray}
Here $k_1$ and $k_2$ are real positive constants which are related to the wall parameters, as we shall see later in Section \ref{Secc:IVA}.
Formally,  Eq. ~\eqref{eq:SchChr} defines the conformal mapping: $z= f(\zeta)$. 

 In the $z$-plane, the complex velocity potential can be written as:
\begin{equation}
w(z)= \phi(z)+i \psi (z)
\end{equation}
and  
\begin{eqnarray}
w(z)=w(f(\zeta)) \equiv \chi(\zeta )=v_0 \zeta=v_0 f^{-1}(z).\label{eq:ComplexVelw}
 \end{eqnarray}

Here $\phi(z)= {\rm Re}(w(z ))$ and  $\psi(z)= {\rm Im}(w(z))$.
Finally, the complex velocity field in the $z$-plane follows by:
\begin{eqnarray}
v= \frac{d w}{dz}=\frac{d w}{d \zeta} \frac{d \zeta} {dz}=\frac{v_0 }{ \frac{d z} {d \zeta}}= \frac{v_0 }{f'(\zeta)} =  v_0 \sqrt{\frac{\zeta^2-k_2^2}{\zeta^2-k_1^2}}, \label{eq:ComplexVelWall}
 \end{eqnarray}
  with $k_1$ and $k_2$ to be determined by the obstacle geometry.

\subsubsection{Relation among the parameters $k_1$ and $k_2$ and the geometrical parameters for the wall: $a$ and $h$.}\label{Secc:IVA}
We now derive the relation between $k_1$ and $k_2$ and the geometrical parameters: $a$ and $h$.
To  achieve this, let us use the formal integration of \eqref{eq:SchChr}, leading to:
\begin{eqnarray}  z(\zeta) = \int_0^\zeta  \sqrt{\frac{u^2-k_1^2}{u^2-k_2^2}} \,du+ z(0) =f(\zeta). \label{eq:SchChrIntegrated1} \end{eqnarray}

Using the fact that $z(k_1)= a+i h$  and $ z(0)= i h$ as shown on Fig.  \ref{Fig:Wall}  and Eq. \eqref{eq:SchChrIntegrated1}, we obtain the value of $a$ and $h$ as function of $k_1$ and $k_2$:
\begin{eqnarray} z(k_1)=a +ih =   \int_0^{k_1}\sqrt{\frac{k_1^2-u^2}{k_2^2-u^2}} \,du+ i h . \label{eq:BCk1}
  \end{eqnarray}

 Therefore the half-width of the wall, $a$, is given by (after a change of variables $u=k_1 \sin t$, see Appendix \ref{App:Parametersk1k2}):

  \begin{eqnarray} 
  \frac{a}{k_2}  &=&    \int_0^{\pi/2 }   \sqrt{1 -\kappa^2 \sin^2  t }   \,  \,dt    -(1-\kappa^2 )  \int_0^{\pi/2 }  \frac{1}{\sqrt{1 -\kappa^2 \sin^2  t }} \,  \,dt =  E\left(\kappa^2\right)   -(1-\kappa^2)K \left(\kappa^2\right) ,  \label{eq:SCWallEqFora}
  \end{eqnarray}
where,    $\kappa = {k_1}/{k_2}<1$, and, 
  $E(z)$ and $K(z)$ are  elliptic functions of first and second kinds~\cite{Arfken}:
  \begin{eqnarray}
K(m)= \int_0^{\pi/2 }\frac{1}{\sqrt{1 -m \sin^2  t } } \,  \,dt   \quad \& \quad 
E(m) = \int_0^{\pi/2 }\sqrt{1 -m \sin^2  t } \,  \,dt  ,   \nonumber  \end{eqnarray}
respectively.
After a straightforward similar calculation (details are given in the Appendix \ref{App:Parametersk1k2}) we obtain: 
\begin{eqnarray}
 \frac{h}{k_2} &= &   \int_{\arcsin(\kappa)}^{\pi/2 }{\sqrt{\sin^2  t -\kappa^2 }} \,  \,dt =  {\rm Im}\left[   \int_{0}^{\pi/2 }{\sqrt{\kappa^2 - \sin^2  t }} \,  \,dt \right] =   {\rm Im}\left[   \kappa \, E\left(1/\kappa^2\right)\right].\label{eq:SCWallEqForh}
  \end{eqnarray}
 
In conclusion, the relations among the parameters $k_1$ and $k_2$ and the geometrical parameters $a$ and $h$  are given by these two Eqs.\eqref{eq:SCWallEqFora} and \eqref{eq:SCWallEqForh}. A particular case is the one of the flow around a square in which case $a=h$, giving $\kappa = 1/\sqrt 2$.

\subsubsection{The fluid velocity}
The fluid velocity is given by  Eq.~(\ref{eq:ComplexVelWall}) in terms of the $\zeta $ variable. On the other hand, the spatial coordinates $z$ are  also given as a function of the $\zeta$ variable by  Eq.~\eqref{eq:SchChrIntegrated1}. Therefore, the equations (\ref{eq:ComplexVelWall}) and \eqref{eq:SchChrIntegrated1} are the formal solution for the velocity field in all space for a given set of parameters $(a,h)$.

Because we are  interested only in the local speed near the corners, that is at $z\approx \pm a+ih$, for $\zeta \approx \pm k_1$ we can derive approximate dependence of the local velocity near the corners. In particular, for $z\approx a+ih$ the complex  velocity Eq.~\eqref{eq:ComplexVelWall} is written in a parametric fashion in terms of the  spatial coordinate $z=x+i y$, leading to:
 \begin{equation} 
 z(\zeta)-( a+ i h ) \approx %\frac{2}{3} \sqrt{\frac{2k_1}{k_1^2-k_2^2}} \left(\zeta-k_1\right)^{3/2} =
  -i\frac{2}{3 } \sqrt{\frac{2k_1}{k_2^2-k_1^2}} \left(\zeta-k_1\right)^{3/2} . %\nonumber
 \label{eq:SchChr4} \end{equation}

Finally, after a straightforward calculation detailed n Appendix  \ref{App:LocalSpeed}, one obtains that the local velocity diverges near the corner as:
 \begin{eqnarray} 
 v (z) \approx   
 v_0 e^{i \pi/3} \left( \frac{1- \kappa^2 }{3  \kappa}   \right)^{1/3}  \left( \frac{k_2}{  z-a- i h }\right)^{1/3} ,\quad {\rm as} \quad  z\to a+i h 
 \label{eq:SchChr5} \end{eqnarray}
 Therefore, within this framework the critical velocity for vortex nucleation should be zero, which is not what is observed in the numerics!
 
\subsubsection{The critical velocity}\label{Secc:CriticalVelocityWall}

In fact, in the  Gross-Pitaevski\u\i~ model, such singular velocity field near the corner is regularized by the quantum pressure that creates a boundary layer inside which the incompressible potential flow is not valid. While this effect was only bringing a small correction for the transonic transition around a cylinder \cite{Frisch1992,Josserand1999,Winiecki1999,HUEPE2000126,Pham2005}, or simpler geometries \cite{Kokubo2025} (where the velocity field remains bounded), it cannot be ignored here.
However, while this problem would require to compute the complex interplay between the quantum pressure and such singular geometries, involving the coupling between the boundary layer, of the order of the healing length and the flow geometry, it is beyond the scope of the present work.
Nevertheless, assuming that the potential velocity is valid until a distance to the wall of the order of the healing length, we can deduce a local velocity that does not diverge but reaches a maximum at  $|z-a- i h| = \alpha \xi_0$, where $\alpha$ is a dimensionless parameter which we shall determine by fitting the numerical data with the prediction of the potential theory. More precisely, our approximation takes advantage that the quantum pressure regularizes the flow near the corner so that the maximum velocity for the transonic transition has to be estimated at a distance of the order of the healing length.
 
 Therefore,  using  Eq. (\ref{eq:criticalvelocitysquare}), we find that  the critical velocity  $v_0^{c}$ obeys the 
 following relation:
 \begin{eqnarray}
  \frac{{v_0^{c}}}{c}   =  \frac{\sqrt{2}}{  \left( \frac{3^{1/3} \left( 1-\kappa^2 \right)^{2/3}  }{  \kappa^{2/3}}\left( \frac{k_2}{ \alpha\xi_0}\right)^{2/3} 
 - 1\right)^{1/2} } , 
\label{eq:criticalvelocitysquarewall}
\end{eqnarray}
in terms  of the parameters $0\leq \kappa\leq1$ and $ {k_2}/{ \xi_0}$. By the use  of Eq. \eqref{eq:SCWallEqFora} and Eq. \eqref{eq:SCWallEqForh} we can deduce  $v_0^{c}$  as a function  of $a$ and $h$  by the following method. For a given $h/\xi_0$, one computes substituting Eq.~\eqref{eq:SCWallEqForh}
into Eq. \eqref{eq:criticalvelocitysquarewall} so that $v_0^{c}$ becomes a function of $h/\xi_0$ and $\kappa$. This relation gives:

 \begin{eqnarray}
  \frac{{v_0^{c}}}{c}   = \frac{\sqrt{2}}{  \left( \frac{3^{1/3} \left( 1-\kappa^2 \right)^{2/3}  }{  \kappa^{2/3}}\left(  \frac{1}{{\rm Im}\left[ \kappa E\left(1/\kappa^2\right)\right] }  \frac{h }{ \alpha\xi_0} \right)^{2/3} 
 - 1\right)^{1/2} } .
\label{eq:criticalvelocityWall}
\end{eqnarray}

The next step  requires  the use the relations  Eqs.~\eqref{eq:SCWallEqFora} and \eqref{eq:SCWallEqForh}
 $$ \frac{a}{  \left(E\left(\kappa^2\right)   -(1-\kappa^2)K\left(\kappa^2\right) \right)}= k_2 = \frac{1}{{\rm Im}\left[ \kappa E\left(1/\kappa^2\right)\right] }  h,  $$
thus,
  \begin{equation} 
  a=  \frac{(E\left(\kappa^2\right)   -(1-\kappa^2)K\left(\kappa^2\right) )}{{\rm Im}\left[ \kappa E\left(1/\kappa^2 \right)\right] }  h .
  \label{eqaplot}
   \end{equation}

 Finally, combining Eqs.~\eqref {eq:criticalvelocityWall} and \eqref{eqaplot} we obtain $v_0^{c}$, as a function of   $a/\xi_0$  by varying the parameters $\kappa$ at fixed $h/\xi_0$. 
 %This dependence is drawn in Fig. \ref{Fig:CriticalMach} (a) and (c) where it is compared with the direct numerical simulations of the GP equation.\\
Lastly, by employing the asymptotic relations (see Eqs.~\eqref{eq:Asymptotic1} and \eqref{eq:Asymptotic2} derived in the Appendix \ref{App:AsymtoticEK}), we deduce asymptotic relations for the critical velocity for the cases of large and small $h/a$ ratio:
 \begin{eqnarray} 
 \frac{v_0^c}{c}  &\approx & \left\{
   \begin{array}{cc} \left(\frac{32 }{3\pi} \right)^{1/6}  \left(\frac{a^{1/6} ( \alpha \xi_0)^{1/3} }{h^{1/2} } \right)& \frac{h}{a} \to \infty\\
     \\
     \left(\frac{\pi^{2}   }{6 }  \right)^{1/6}  \left(  \frac{  \alpha \xi_0  } { \, h } \right)^{1/3} & \frac{h}{a}\to0\quad \& \quad \frac{\xi_0}{h} \ll 1
      \end{array} 
       \right. \, .
       \label{eq:McAsympWall}
\end{eqnarray}
The specific case where $a/\xi_0$ will be discussed further on below.

\subsection{Incompressible flow  passing around a well domain}\label{Secc:CriticalVelocityWell}
\begin{figure}[h]
\begin{center}
\centerline{\includegraphics[width=10cm]{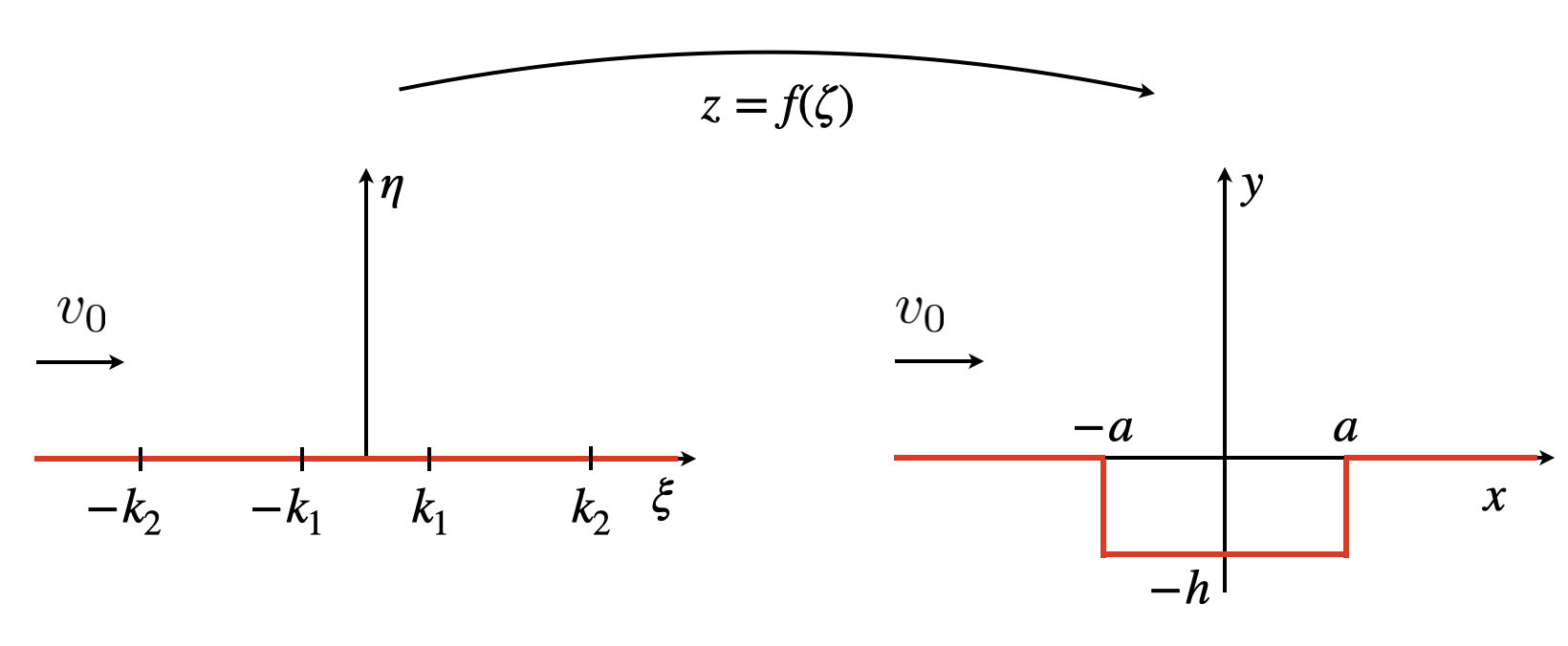} }
\caption{ \label{Fig:well} 
Conformal map for transforming the plane $\zeta$-plane into a well in the $z$-plane. Here  $f(0)= -i h$, $f(\pm k_1) =\pm a-i h$, $f(\pm k_2)= \pm a$.
}
\end{center}
\end{figure}

Now, we consider the case of a well domain, see Fig. \ref{Fig:well}. As expected, this case can also be solved exactly using complex variables, following the same protocol than for the wall.
There, the Schwarz-Christoffel transform reads:
\begin{eqnarray}  \frac{dz}{d\zeta} = \sqrt{\frac{\zeta^2-k_2^2}{\zeta^2-k_1^2}} .   \label{eq:SchChrForWell}  \end{eqnarray}
Now the constants $k_1$ and $k_2$ follow by the boundary conditions $z(0)= -i h$, $z(k_1) =a-i h$ and $z(k_2)= a$ as shown on Fig.  \ref{Fig:well}. 

Similarly, the parameters $k_1$ and $k_2$ are given by relations involving elliptic functions:
   \begin{eqnarray} \frac{a}{k_2} =   \int_0^{\pi/2 }{\sqrt{1 -\kappa^2 \sin^2  t }} \,  \,dt = E\left(\kappa^2\right)  \quad & \& & \quad 
  \frac{h}{k_2} =\int_{\arcsin(\kappa)}^{\pi/2 }\frac{\cos^2  t}{\sqrt{\sin^2  t -\kappa^2 }}    \,dt.
   \label{eq:BCWellFora&h}
  \end{eqnarray}
  
 Then, we find that the critical velocity (all calculations are done explicitly in the Appendix \ref{App:TheCaseWell}), reads:
 \begin{eqnarray}
  \frac{{v_0^{c}}}{c}   =  \frac{\sqrt{2}}{  \left( {3^{1/3} \left( 1-\kappa^2 \right)^{2/3}  }\left( \frac{k_2}{  \alpha\xi_0}\right)^{2/3} 
 - 1\right)^{1/2} } .
\label{eq:criticalvelocityWell}
\end{eqnarray}
In the current case, $\kappa=k_1/k_2<1$, and the constants $k_1$ and $k_2$ are given by Eqs.~\eqref{eq:BCWellFora&h}.

Finally, we provide also asymptotic scaling laws for the case of a deep well ($h/a\gg 1$),  and the opposite limit ($h/a\ll 1$):
%The critical velocity  Eq.~\eqref{eq:criticalvelocityWell} reads (The asymptotic relations are detailed in the Appendix \ref{App:TheCaseWell}.):
  \begin{eqnarray}
  \frac{v_0^{c}}{c}   =\left\{
  \begin{array}{cc}
   \frac{\sqrt 2}{  \left( {3^{1/3}  }\left(   \frac{2a }{\pi \alpha\xi_0}  \right)^{2/3} 
 - 1\right) ^{1/2}}\approx  \left( \frac{2\pi^2}{ 3  }\right) ^{1/6} \left(   \frac{ \alpha\xi_0}  {a }\right)^{1/3} 
 , & {\rm for}\quad h/a \to \infty\\
  \frac{\sqrt 2}{  \left( {2\times 6^{1/3}  }\left(   \frac{h }{\pi  \alpha\xi_0}  \right)^{2/3} 
 - 1\right)^{1/2}  } \approx  \left(\frac{\pi^{2}   }{6 }  \right)^{1/6}   \left(\frac{ \alpha\xi_0}{h} \right)^{1/3}  , &   {\rm for}\quad h/a \to 0
 \end{array}
  \right. .
\label{eq:McAsympWell}
\end{eqnarray}
Notice that, because of the obvious asymmetry between the two cases, these scaling laws are different from those obtained for the wall obstacle. 
%As we shall see in Section~\ref{Sec:Results},  the limit  for  the wall and  well  critical velocity   for  $h/a \to 0$  which  are  given respectively by  Eq.~\eqref{eq:McAsympWall}  and  Eq.~\eqref{eq:McAsympWell} are found  in good agreement with  those  obtained by our  numerical simulations.

\subsection{Special limits}

In the previous analysis, we considered implicitly that the thickness of the wall $a$ was larger than the healing length $\xi_0$, while we estimate the critical velocity inside this boundary layer. Similarly, a step can be modeled as the side of a well in the limit  $a \to\infty$. We will investigate these two specific limits below.

\subsubsection{A thin barrier: the limit $a\to0$.}
At first glance, Eq.~\eqref{eq:McAsympWall} implies that in the limit $a\to0$, the critical velocity vanishes which is not physical. Indeed when $a\to 0$  the finite-width wall becomes an infinitely thin barrier with an external  turning angle of $ \pm \pi$ as in Refs. \cite{rica1993critical,rica2001vortex,Kokubo2024,Kokubo2025}. 

This specific limit must be taken as limit $k_1\to 0 $ in Eq.~\eqref{eq:ComplexVelWall}, that is 
$ \frac{dz}{d\zeta} =  \frac{ \zeta}{\sqrt{\zeta^2-k_2^2}}$, hence, after a direct integration $ z =  \sqrt{\zeta^2-k_2^2} . $
Identifying $\zeta=0$ with $z= ih$, one sets $k_2=  h$.
Therefore, the complex potential velocity reads 
$$f(z) = v_0 \sqrt{ h^2 + z^2}.$$
Finally, after computing the local velocity near $z\approx i h$ and using the criterion for the  critical velocity Eq.~\eqref{eq:criticalvelocitysquare}, one obtains \cite{rica2001vortex}:
$$\frac{v_0^{(c)}}{c} \sim\frac{2}{\sqrt{3}} \left(\frac{\alpha\xi_0}{h}\right)^{1/2}.$$
The matching of this critical velocity with Eq.~\eqref{eq:McAsympWall} considering that the limit $a\to 0$, formally corresponds to $a \to \frac{2\pi}{9} \alpha \xi_0$, because of the presence of the boundary layer at the barrier.

\subsubsection{A step: the limit $a\to\infty$.}

Formally speaking, this limit $a\to\infty$ corresponds to the case of a step, in which case the
 conformal map reads \cite{milne1996theoretical}:
$$ \frac{dz}{d\zeta} = \frac{1}{v_0} \sqrt{\frac{\zeta+\frac{2 v_0 h}{\pi} }{\zeta}},$$
hence, the fluid velocity in terms of $\zeta$ is
$
v(z)=
v_0\sqrt{\frac{\zeta}{\zeta+\frac{2 v_0 h}{\pi} }} .
$

The corner of interest  corresponds to $\zeta\approx-\frac{2 v_0 h}{\pi}.$ By integrating explicitly the conformal map near the region of interest, one obtains 
$ z \approx ih - \frac{ i \sqrt{2\pi}}{3 {v_0}^{3/2} h^{1/2} } \left(\zeta+\frac{2 v_0 h}{\pi}\right)^{3/2}.$ Therefore, 
the local speed becomes 
$|v(z)| \approx v_0  \frac{ 2^{2/3} h^{1/3}}{  (3\pi)^{1/3} } |{z-i h} |^{-1/3} , $
and, finally,  the  critical  velocity becomes:
\begin{equation}
\frac{v_0^c}{c} =  \left(\frac{\pi^{2}   }{6 }  \right)^{1/6}   \left(\frac{\alpha\xi_0}{h} \right)^{1/3} \,  ,
\label{eq:criticalvelocitysquareStep}
\end{equation}
in agreement with the critical velocity  given in Eq.~\eqref{eq:McAsympWell}.

 \section{Comparisons of the results of the numerical simulations and  the theory}\label{Sec:Results}
 
 In this section,  we compare the results obtained by the numerical simulations done using the protocol described in Section \ref{Secc:Numerics}. Initially, we study the effect of the system size on the critical velocity, and then after obtaining a satisfactory and reproducible domain size, we perform a systematic study of the critical velocity as a  function of the obstacle parameters that we compare with the theoretical prediction computed in previous Section \ref{Sec:CriticalVelocity}. 

 \subsection{ Finite size effect }
 
  Our numerical simulations   are  based on a  finite system size, $L_x$ and $L_y$, while our  analytical theory is developed for infinite space for simplicity.  We  therefore need to estimate when finite size effects are relevant. A first condition was already notice in  \ref{subSecc:Obstacles}, because of periodic boundary conditions the half width  $a$ of the obstacle must be lower than  $L_x/4$. For $a>L_x/4$ a wall becomes a well.  This fact has the convenience that we do not need to simulate separately the wall and the well, since the latter is a straight forward continuation of the former, and vice-versa.

\begin{figure}[h]
\begin{center}
\centerline{\includegraphics[height=5cm]{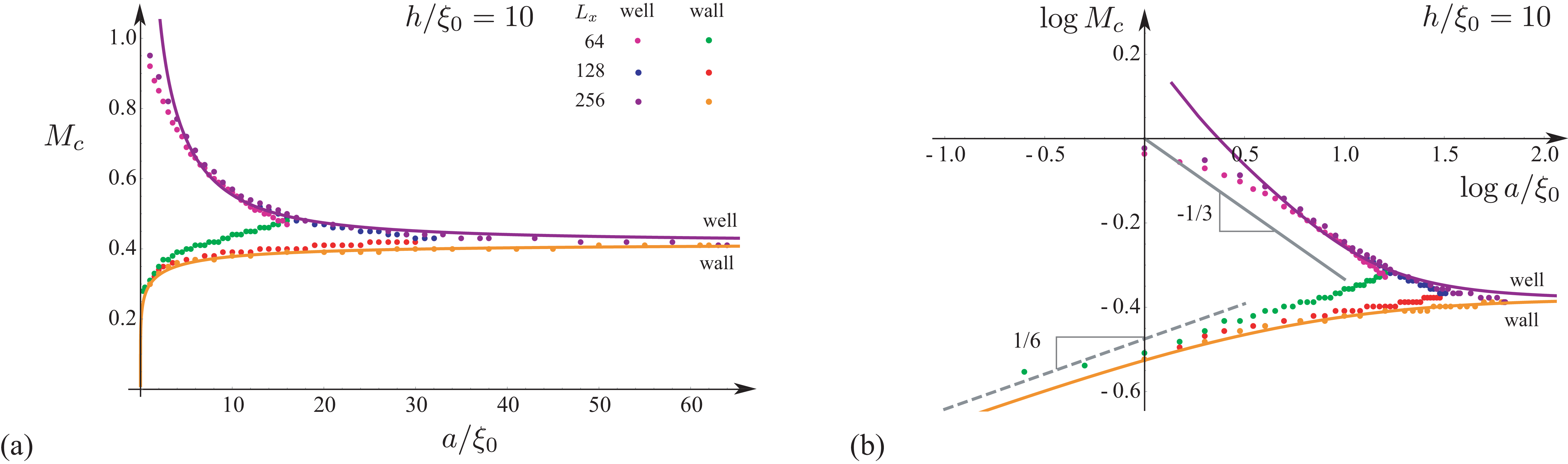} }
%\centerline{(a)\includegraphics[height=4.5cm]{Fig5ah15.eps} \quad (b)\includegraphics[height=4.5cm]{Fig5bh15.eps} }
\caption{ \label{Fig:Fig5} 
 Critical Mach number as a function of the width $a/\xi_0$ of the wall (or well) obstacle. The numerical simulations are for $h/\xi_0 =10$, $|\psi_0|^2=1$, $\xi_0=1$, $dx= 0.5$ and $dt=1/100$. 
The dots correspond to the numerical simulations for  the wall and the well  for different system size: $L_x = 64$  (green: wall, magenta: well),  $L_x = 128$  (red: wall, blue: well),  $L_x = 256$  (orange: wall, purple: well).  The continuous lines correspond to the theoretical calculations using Schwarz-Christoffel mapping for $h/\xi_0 =10$ and the adjustable parameter $\alpha \simeq 0.5$ for the wall (orange line), and the well (purple line). (See Table \ref{tab:table1}). (a) Plot in linear scale. One readily notices an already strong asymmetry for $a< h$. (b) Plot in log-log scale (in base 10). As a guide for the eye, we plot with a segmented gray line a slope $1/6$ for comparisons with the scaling laws given by Eq.~\eqref{eq:McAsympWall} for $h\gg a$,  and  we plot with a continuous gray line a slope $-1/3$ for comparisons with the scaling laws given by Eq.~\eqref{eq:McAsympWell} for $h\gg a$.}
\end{center}
\end{figure}

In Fig. \ref{Fig:Fig5} we compare the result of numerical simulations for  $h/\xi_0=10$ for three different system sizes: $L_x= 64,128 \,\&\, 256$ units of $\xi_0$, with the critical velocity given by the theory of compressible flow in infinite space (Eqs.~\eqref{eq:criticalvelocityWall} and \eqref{eq:criticalvelocityWell}). One notices that the theory agrees well with the numerics for  $L_x= 128 \,\&\, 256$. The agreement is realized by fitting the dimensionless parameter $\alpha$ used in the final expression Eq.~\eqref{eq:criticalvelocityWall} for the flow around a wall. For $h/\xi_0=10$, $L_x= 128 \,\&\, 256$ the fit gives $\alpha \approx 0.5$. Importantly, the value $\alpha \approx 0.5$ obtained to fit the data of the flow around a wall works also for the case of the flow over a well. Additionally, Fig \ref{Fig:Fig5}(b) compares favorably the numerical calculations with  the asymptotic behavior  $v_0^c \sim a^{1/6} $ of Eq.~\eqref{eq:McAsympWall} for large $h/a$, and a full gray line a slope of -1/3 for comparison with the  scaling law of Eq.~\eqref{eq:McAsympWall}  for large $h/a$.

In conclusion, we obtain that the system sizes: $L_x= 128$ and $L_x=256$, give essentially  similar results. However,  for  $L_x= 128$ and larger values of $h>25\xi_0$ there are some slight deviations due to the finite system size.  Therefore for the rest of the article we will use $L_x=L_y=256$,
$dx=0.5$ and $dt=0.01$. 

  For the case $L_x=64$ typical  finite size effects are observed. Here, a possible improvement for the comparison between the theory and such numerical simulations would be the use of a periodic Schwarz-Christoffel mapping (see  for example Ref.~\cite{Baddoo2019}), as well as, the presence of the Neumann boundary conditions at the vertical domain $y=L_y$, and, lastly,  the effect of compressibility which may be treated by a Janzen-Rayleigh perturbation \cite{Rica2001,Kokubo2025}. Nevertheless, all these variations are beyond the scope of the present study.

\subsection{Dependence  of $v_0^{(c)}$ as a function of the width $a$ and of the height/depth $h$ of the wall/well}

Thus, for $L_x=L_y=256$, corresponding to a numerical lattice of $N_x\times N_y=512 \times 512$, since the mesh size is $dx=0.5$, the numerical simulations of the G-P equations are found in excellent agreement with the theoretical predictions both for the cases of the wall and the well. Furthermore, we observe a clear difference between the barrier and the well as the half width $a$ varies: the critical velocity increases when the size $a$ of the wall (barrier) increases, while the opposite holds  for the well, because of the geometrical asymmetry between the wall and the well.  

\begin{figure}[h]
\begin{center}
\centerline{\includegraphics[height=10cm]{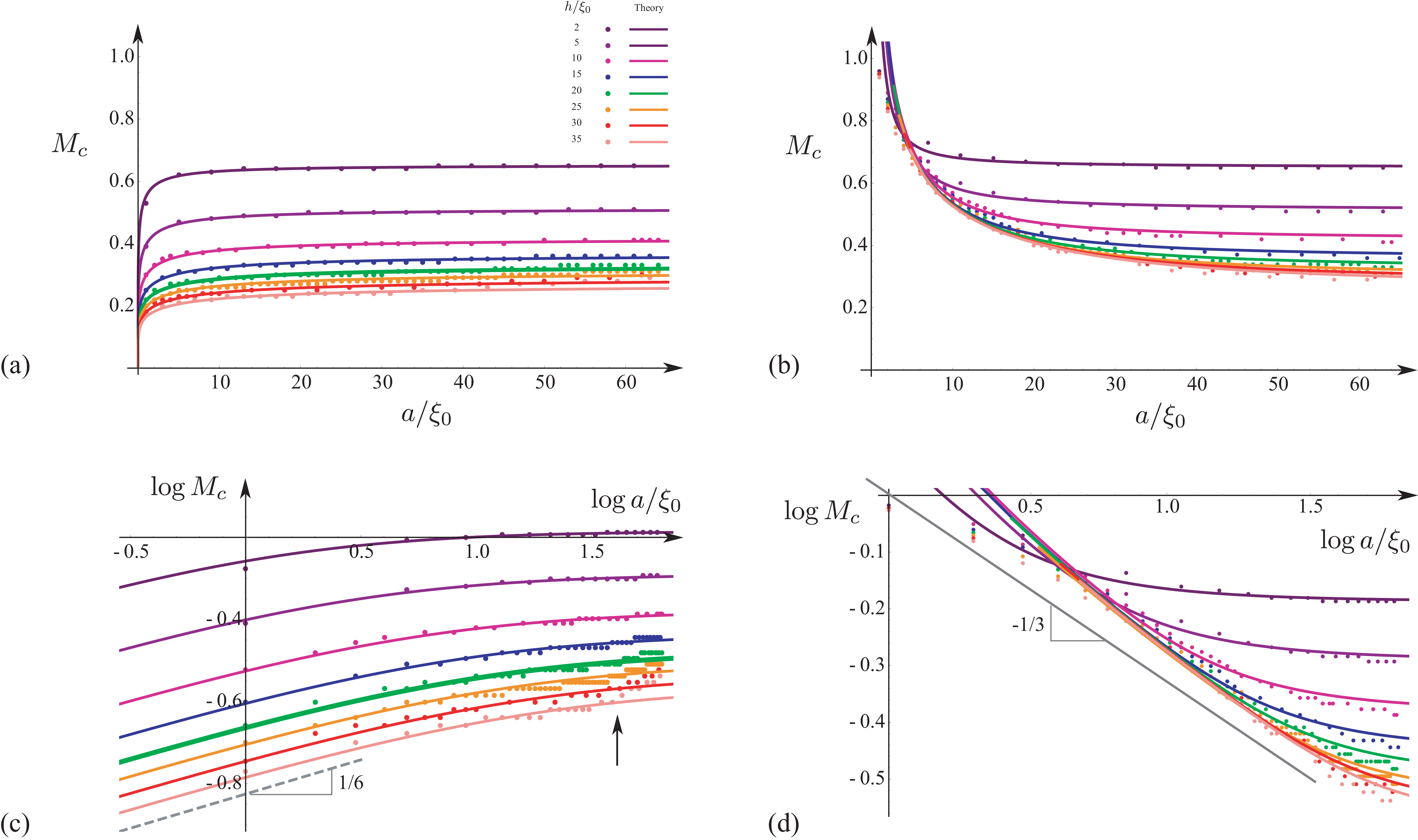} }
\caption{ \label{Fig:CriticalMach} Critical Mach number for a wall (in linear scale (a) and in log-log scale (c)) and well (in linear scale (b) and in log-log scale (d)) type of obstacle. The numerical simulations are for $|\psi_0|^2=1$, $\xi_0=1$, $L_x=L_y=256$, $dx= 0.5$, $dt=1/100$ and for $h/\xi_0 =\left\{  2,5,10,15,20,25,30, 35 \right\}$. 
 The dots correspond to the numerical simulations for  $h/\xi_0 =2$ (dark purple) $h/\xi_0 =5$ (purple), $h/\xi_0 =10$ (violet), $h/\xi_0 =15$ (blue),  $h/\xi_0 =20$ (green), $h/\xi_0 =25$ (orange), $h/\xi_0 =30$ (red), and $h/\xi_0 =35$ (pink), as labeled in panel (a). The continuous lines correspond to the theoretical calculations using  Eq.~\eqref{eq:criticalvelocityWall} for the wall and Eq.~ \eqref{eq:criticalvelocityWell} for the well, using  for each $h/\xi_0$, a parameter $\alpha$ as in Table \ref{tab:table1}.  Both theoretical predictions are drawn with the same colors as in panel (a). The arrow in panel (c) indicates the region of discrepancy between the theory and the numerics for $h\gtrsim 30 \xi_0$ due to finite size effects.}
\end{center}
\end{figure}

More quantitatively,  the results of our numerical simulations are confronted on Fig. \ref{Fig:CriticalMach} with the predicted theoretical critical velocity given by  Eqs.~\eqref{eq:criticalvelocityWall} and \eqref{eq:criticalvelocityWell} in Sections \ref{Secc:CriticalVelocityWall} and \ref{Secc:CriticalVelocityWell} respectively.  We observe that $M_c$ always decreases as $h$ increases for a fixed value of $a$ for both cases. Moreover, $M_c$ increases as $a$ increases, at least for $a\gtrsim\xi_0$, for a fixed value $h$ (see Fig. \ref{Fig:CriticalMach}-(a)), while the opposite happens for the well Fig. \ref{Fig:CriticalMach}-(b). Finally, in  Fig.
 \ref{Fig:CriticalMach} (c) \& (d)  we plot the critical velocity in log-log scale and compare the asymptotic behavior for the limiting case $h/a \rightarrow \infty$  for the wall $v_0^c\sim a^{1/6}$  as given by Eq.~\eqref{eq:McAsympWall} and $v_0^c\sim a^{-1/3}$ for the well as given by Eq.~\eqref{eq:McAsympWell}. 

As mentioned above these differences comes from the asymmetry between the wall and the well. Thus, even though the exterior angles of the wall  and the well edges are  the same ($3\pi/2$), the critical velocity for vortex nucleation is found higher for the well than for the  wall.  However, on the one side, a wall is a  concave-like perturbation which impedes the mean flow and repels the streamlines from the boundaries in the open semi-infinite system, compacting the streamlines with the consequence of increasing the local speed near the obstacle. On the other side, the well is a convex-like perturbation that creates a dilation of the streamlines in the convex cavity, decreasing the local speed. 
As the width $2a$ increases, the action by the well  on the streamlines decreases, and, thus, the critical velocity decreases, while for the wall the `convexity' decreases, smoothing the effect of the wall, increasing the critical velocity. Eventually, in the limit $a\to\infty$ the asymmetry for the critical velocity disappears as observed in Figs. \ref{Fig:Fig5}  and \ref{Fig:CriticalMach}. 

Qualitatively, for both the wall and the well the vortices are nucleated close to the sharp edges where the velocity is greater, and are advected by the flow. 
Just  above the critical  velocity the vortices tend to follow the frontiers of the system. Thus, in the case of the wall the vortices move in a semi-infinite open  space  system  above the wall, while for the well the vortices tend to move inside  the cavity and then have to develop in a confined zone between the two boundaries of the well  impeding  vortex formation (See Fig. \ref{Fig:NumericsWall} (c)). 

We have noticed in Fig. \ref{Fig:CriticalMach} that small deviations between the theory and the numerical simulations occur for large value of $h\gtrsim 30 \xi_0 $. These small deviations  appear  for a value approaching  $a=L_x/4$ and can be interpreted as finite size effect mayin both  $x$ and $y$ directions (indeed, when $a=L_x/4$  a wall and a well are  indistinguishable objects due to the periodicity of the system). Finite size effects can be diminished as we have shown on Fig. \ref{Fig:Fig5} in which we have simulated a larger system.  As already  mentioned, a more complete predictive theory here would require to use complex periodic Schwarz-Christoffel mapping as used in \cite{Baddoo2019}.

Finally, we end this section by commenting on the parameter $\alpha $ introduced in Secc. \ref{Secc:CriticalVelocityWall}. This parameter $\alpha$ includes the missing effects in the theory based  on the interplay between compressible flows around  obstacles and the  effect of the quantum pressure which are dispersive effects  preventing shock formation. At a distance  of the order $\alpha \xi_0$, that is inside the healing boundary layer, the potential, irrotational, and compressible flow is not longer valid. In the present paper we simply fitted the parameter $\alpha$, as 
 the parameter $h/\xi_0$ varies by using the numerical findings obtained in Fig.~\ref{Fig:CriticalMach} leading to a very good agreement between the theory and the simulations. 
The values of $\alpha$ which allow for such very good comparison between the theory and the numerical simulations are shown in Table \ref{tab:table1} as $h/\xi_0$. We notice that $\alpha$ increases weakly as $h$ increases, but we expect that $\alpha$ saturates to a fixed value of the order of $\alpha \approx 0.52$. 

The possible determination of the  value of $\alpha$ is still an open question which will be investigated in the future. 
In principle, the dimensionless parameter $\alpha$ may be computed by matching the inner asymptotic zone, which is a linear Schr\"odinger equation with the outer compressible potential irrotational flow. However, these asymptotic calculations are out of the scope of this article and require advanced matching techniques which take into account point like divergence due to  the presence of sharp edges effect.

  \begin{table}[h!]
  \begin{center}
    \begin{tabular}{|c|c|c|} \hline\hline
     $h/\xi_0$ & $\alpha_{\rm wall} $  &  $\alpha_{\rm well} $ \\ \hline\hline
     2&0.324 & 0.323 \\ \hline
 5 & 0.436 &  0.438 \\ \hline
 10 & 0.498 & 0.501 \\ \hline
 15 &  0.520 & 0.483 \\ \hline
20 & 0.536  & 0.478 \\ \hline   
25 & 0.549 &0.470 \\ \hline
30 &0.543&0.474  \\ \hline
35 & 0.519&0.470
   \\ \hline
      \hline
          \end{tabular}
              \caption{Numerical parameters used in the simulation for $L_x=L_y=256$ units of $\xi_0$.  For some values there is a 10\% of difference between the values estimated for a wall, $\alpha_{\rm wall}$, and a well $ \alpha_{\rm well} $. For $h/\xi_0=10$, one notices that $\alpha\approx 0.5$ as used in Fig. \ref{Fig:Fig5}.}
    \label{tab:table1}
  \end{center}
\end{table}

\section{ Conclusion and perspective}\label{Sec:Conclusion}

In this work, we have investigated analytically and numerically the transition from a free vortex flow to a vortical flow in a model of superflow for different geometries. We have used the Gross-Pitaevski\u\i~ equation which is a model equation for superflow that captures the transitions between a potential flow and a vortex-laden irreversible flow when the speed exceeds a geometry-dependent critical threshold. We have simulated a superflow around specific sharp-angle geometries such as the flows over a wall or a well of finite size by gradually ramping up the flow speed until observing the onset of vortex nucleation.
The results of our numerical simulations of the full Gross-Pitaevski\u\i~ equation in two space dimensions are found in very good agreement with our analytical model. The theoretical framework of our model consists in computing the potential flow around rectangular geometries using conformal mapping techniques. More specifically, we employ the Schwarz-Christoffel transformation to map the exterior of a rectangular barrier to the upper half complex plane, allowing the construction of analytic expressions for the velocity field in the ideal (incompressible and irrotational) limit. This method enables a precise calculation of the regions of maximum flow velocity, which play a pivotal role in determining where and when vortex nucleation may occur. To estimate the critical velocity, we have applied the criterion introduced in Refs.~\cite{Frisch1992,Josserand1999}, which asserts that vortex nucleation begins when the local fluid velocity reaches a threshold for the transonic transition. Mathematically, this means that the elliptic equation for the velocity potential becomes hyperbolic as the flow speed increases \cite{Landau1987Fluid}. By combining the analytic velocity field obtained from potential theory with this criterion, we have derived critical velocities for vortex formation for a flow passing a rectangular wall or a rectangular well. We have found that the critical velocity scales asymptotically as power laws before reaching a plateau as a function of the system parameters, which are the barrier height $h$ and its half-width $a$.
Moreover, we have found that the critical velocity for vortex nucleation around a wall increases as its width increases, and the converse holds for the well.
%Our theoretical predictions for vortex nucleation based on a complex mapping model (Schwarz–Christoffel mapping) for the potential flow are in excellent agreement with our numerical simulations of the Gross-Pitaevski\u\i~ equation, since they reproduce with good accuracy the value of the theoretical velocity for vortex nucleation.
We have noticed that for small obstacle width, small $a$ values in comparison to the healing length value $\xi_0$, particularly for walls, the predictions from potential theory tend to slightly overestimate the critical velocity found numerically. This discrepancy arises due to compressibility effects, which become increasingly significant at smaller scales comparable to the healing length while they are neglected in the incompressible framework of the potential theory. We believe that the comparison between the numerical values and the analytical prediction for small $a$ and also small $h$ could be improved by taking into account the compressibility of the fluid via the Janzen–Rayleigh expansion, as done in \cite{Rica2001,Kokubo2025}. Another small discrepancy between the theory and the numerical results arises due to finite-size effects and periodic boundary conditions, which are mostly visible for large obstacle height $h$, and could be investigated by using complex periodic Schwarz–Christoffel mapping analytical techniques as in Ref.~\cite{Baddoo2019}.

Overall, our study highlights the versatility of conformal mapping techniques in analyzing complex superfluid flows and extends the theoretical understanding of vortex nucleation beyond the canonical circular or elliptical obstacle configurations. By bridging analytic methods and numerical simulations, we have elucidated how geometry governs the critical behavior in superfluid hydrodynamics, with potential implications for the design of quantum microfluidic devices and for the future interpretation of experimental data in superfluid helium, atomic condensates, and nonlinear optical systems with fluids of light. Our numerical simulations of the Gross-Pitaevski\u\i~ equation, which intrinsically incorporate the compressibility of the fluid and nonlinear dynamical effects, will allow future studies aimed  at controling vortex formation.

\section*{Acknowledgments}
\noindent
T.F.  acknowledges fruitful discussions with M\'ed\'eric Argentina and Sergey Nazarenko and the support from the Ecole Universitaire de Recherche SPECTRUM–RISE program (Universit\'e C\^ote d'Azur) and from FONDECYT (Chile) under Grant No.~1220369.
S.R. acknowledges support from the Ecole Universitaire de Recherche SPECTRUM
(Universit\'e C\^ote d'Azur) and from FONDECYT (Chile) under Grant No.~1220369.

\section*{Appendix }\label{Appendix}

\subsection{Conformal mapping of a wall geometry}

\subsubsection{Parameters of the conformal mapping in terms of geometry: the case of a wall.}\label{App:Parametersk1k2}

After the change of variables $u = k_1 \sin  t$, $ {du} = k_1 \cos t \, {dt}$,
 from Eq.~\eqref{eq:BCk1}, we obtain: 
\begin{eqnarray} a =   \int_0^{\pi/2}\sqrt{\frac{k_1^2-k_1^2 \sin^2  t}{k_2^2 -k_1^2 \sin^2  t }} \, k_1 \cos t \,  \,dt =    \frac{k_1^2 }{ k_2} \int_0^{\pi/2 }\frac{\cos^2  t}{\sqrt{1 -\kappa^2 \sin^2  t }} \,  \,dt, =  k_2 \kappa^2 \int_0^{\pi/2 }\frac{\cos^2  t}{\sqrt{1 -\kappa^2 \sin^2  t }} \,  \,dt,  \nonumber
  \end{eqnarray}
  where $\kappa = {k_1}/{k_2}<1.$
Thus, $a/k_2$ is given by Eq.~\eqref{eq:SCWallEqFora}.

   In order to find  $h$ we do similarly, by noticing that
  for $k_1<\zeta<k_2$, thus it is convenient to integrate  Eq.~\eqref{eq:SchChr} in the following way:
  \begin{eqnarray}  z(\zeta)  =   z(k_1)  - i\int_{k_1}^\zeta  \sqrt{\frac{u^2-k_1^2}{k_2^2 -u^2}}\, du . \label{eq:SchChrIntegrated2} \end{eqnarray}
 As shown on Fig. \ref{Fig:Wall}, at $\zeta=k_2$   we have  $z(k_2)= a $ and since $z(k_1)=a+i h$, we conclude:
  \begin{eqnarray} z(k_2)=  a = a+i h  -i \int_{k_1}^{k_2} \sqrt{\frac{u^2-k_1^2}{k_2^2 -u^2}}\, du, \quad \Rightarrow \quad h=   \int_{k_1}^{k_2} \sqrt{\frac{u^2-k_1^2}{k_2^2 -u^2}} . \nonumber
  \end{eqnarray}

After a new change of variables $u = k_2 \sin  t$, $ {du} = k_2 \cos t \, {dt}$
$$ h =   \int_{\arcsin(k_1/k_2)}^{\pi/2}\sqrt{\frac{k_2^2 \sin^2  t-k_1^2}{k_2^2 -k_2^2 \sin^2  t }} \, k_2 \cos t \,  \,dt.
$$
From this expression we obtain Eq.~\eqref{eq:SCWallEqForh}.

\subsubsection{Local speed near the $3\pi/2$ corners}\label{App:LocalSpeed}

We are interested in the local speed near the corners $z\approx \pm a+ih$, that is for $\zeta \approx \pm k_1$. In particular, from  Eq.~\eqref{eq:ComplexVelWall} for $z\approx a+ih$ the local velocity reads: 

 \begin{equation} 
v \approx  v_0\sqrt{\frac{k_1^2-k_2^2}{2k_1}} \frac{1}{\left(\zeta-k_1\right)^{1/2}}= i \, v_0\sqrt{\frac{k_2^2-k_1^2}{2k_1}} \frac{1}{\left(\zeta-k_1\right)^{1/2}}, 
 \label{eq:SchChr2} 
 \end{equation}

 In order to recover the velocity as a function of the spatial coordinates in $z$-plane, one has to use the Schwarz-Christoffel transformation, from Eq.~\eqref{eq:SchChr}, locally around the corner $z \approx a+i h$ ($\zeta \approx k_1$)
$$
  \frac{dz}{d\zeta} \approx \sqrt{\frac{2k_1}{k_1^2-k_2^2}} \left(\zeta-k_1\right)^{1/2} 
$$
thus:
  $$
 z(\zeta)-( a+ i h ) \approx \frac{2}{3} \sqrt{\frac{2k_1}{k_1^2-k_2^2}} \left(\zeta-k_1\right)^{3/2} = -i\frac{2}{3 } \sqrt{\frac{2k_1}{k_2^2-k_1^2}} \left(\zeta-k_1\right)^{3/2} .
 $$
 In the last equation,  we choose the negative value of the square root since the imaginary part of $z$ decreases as $\zeta$ increases for $k_1 \leq \zeta \leq k_2$  as  shown in Fig. \ref{Fig:Wall}.
 Thus, from the above formula:
  \begin{equation} 
   \left(\zeta-k_1\right)^{-1/2}=\left(\frac{ - 2 i \sqrt{2 k_1}} {3 (z-a-ih)   \sqrt{k_2^2-k_1^2}}\right)^{1/3}
   \label{eq:local}
 \end{equation} 
 Writing
 Eq.~\eqref{eq:SchChr2} in terms of $(z-a-ih) $, one obtains the final expression Eq.~\eqref{eq:SchChr5}.

\subsubsection{Asymptotic behaviors of elliptic functions for the case of a wall.}\label{App:AsymtoticEK}
The asymptotic behavior are:
  \begin{eqnarray} 
 \frac{a}{k_2}  =E\left(\kappa^2\right)   -(1-\kappa^2)K(\kappa)&\approx & \left\{
   \begin{array}{cc}
    \frac{\pi}{4} \kappa^2 + \dots & \kappa \to 0 \\
    \\
    1-\frac{1}{2} (1-\kappa ) \left(\log \left(\frac{8}{1-\kappa}\right)+1\right)+\dots & \kappa\to1
      \end{array} 
       \right.
       \quad ,
         \label{eq:Asymptotic1}
     \end{eqnarray}
     and 
      \begin{eqnarray} 
 \frac{h}{k_2}   =  {\rm Im} \left[\kappa E\left(1/\kappa^2\right)\right] &\approx & \left\{
   \begin{array}{cc}  {\rm Im} \left[  i+ \frac{\pi}{4} \kappa ^2 +
     i \frac{\kappa^2}{2}   \log (\kappa/4)  - i\frac{\kappa^2}{4}   +O\left(\kappa ^3\right) \right] & \kappa \to 0\\
     \\
    {\rm Im} \left[ 1-\frac{1}{2}  \left(1+\log 8  - \log \left( e^{i\pi} (1-\kappa )\right) \right) (1-\kappa )+O\left((\kappa -1)^2\right)+\dots\right] & \kappa\to1
      \end{array} 
       \right. \, ,\nonumber\\
       &\approx & \left\{
   \begin{array}{cc} 1 +
     \frac{\kappa^2}{2}   \log (\kappa/4)  - \frac{\kappa^2}{4}   +O\left(\kappa ^3\right)  & \kappa \to 0\\
     \\
   \frac{\pi}{2} (1-\kappa )+O\left((1-\kappa )^2\right)& \kappa\to1
      \end{array} 
       \right. \, .
          \label{eq:Asymptotic2}
     \end{eqnarray}

Lastly, the ratio becomes
     \begin{eqnarray} 
   \frac{h}{a}  &= & \left\{
   \begin{array}{cc} \frac{ 1 +
     \frac{\kappa^2}{2}   \log (\kappa/4)  - \frac{\kappa^2}{4}   +O\left(\kappa ^3\right) }{  \frac{\pi}{4} \kappa ^2 
      +O\left(\kappa ^3\right) } \approx  \frac{4}{\pi \kappa^2 } + \dots  & \kappa \to 0\\
     \\
 \frac{  \frac{\pi}{2} (1-\kappa )+O\left((1-\kappa )^2\right)}{ 1-\frac{1}{2}  \left(1+\log 8  - \log  \left(1-\kappa \right) \right) (1-\kappa )+O\left((\kappa -1)^2\right)} \approx  \frac{\pi}{2} (1-\kappa )  + \dots & \kappa\to1
      \end{array} 
       \right. \, .
  \end{eqnarray}

\subsection{The case of a well domain } \label{App:TheCaseWell}

\subsubsection{Local speed near the $3\pi /2$ corners}

The local velocity reads,
 \begin{eqnarray} 
 v = v_0 \frac{d\zeta}{dz} \quad {\rm with}\quad \frac{d\zeta}{dz} =    \sqrt{\frac{k_2^2-k_1^2}{2k_2}}  \frac{1}{\left(\zeta-k_2\right)^{1/2}},
 \label{eq:SchChr1Well} \end{eqnarray}
   and the velocity in the $\zeta$-plane can be approximated as:
$$
  v \approx {v_0} \sqrt{\frac{2k_2}{k_2^2-k_1^2}} \left(\zeta-k_2\right)^{1/2} .
$$
After an integration of Eq. \eqref{eq:SchChr1Well} one obtains:
$$
 z(\zeta)- a \approx \frac{2}{3} \sqrt{\frac{2k_2}{k_2^2-k_1^2}} \left(\zeta-k_2\right)^{3/2}  .$$
Therefore, one obtains the local velocity of the order of:
 \begin{eqnarray} 
 v (z) \approx v_0 \frac{ \left( k_2^2 -k_1^2\right)^{1/3}  }{3^{1/3} k_2^{1/3} \left(z-a \right)^{1/3} } =  v_0 \frac{ \left( 1-\kappa^2 \right)^{1/3}  }{3^{1/3} }\left( \frac{k_2}{z-a}\right)^{1/3} .
 \label{eq:SchChr2Well} \end{eqnarray}

The condition for the critical velocity follows from Eq,~\eqref{eq:criticalvelocitysquare} and leads to  a critical velocity given by Eq.~\eqref{eq:criticalvelocityWell}.

  \subsubsection{Asymptotic behaviors of elliptic functions for the case of a well.}

In the case of a well obstacle, the asymptotic behavior are:
  \begin{eqnarray} 
 \frac{a }{k_2}  =E\left(\kappa^2\right) &\approx & \left\{
   \begin{array}{cc}
   \frac{\pi }{2}-\frac{\pi  \kappa ^2}{8}-\frac{3 \pi  \kappa ^4}{128}+\dots  & \kappa \to 0\\
    \\
     1- \frac{1}{2} (1-\kappa ) [\log (1-\kappa )+1-3 \log 2]+\dots   & \kappa\to1
      \end{array} 
       \right.
       \quad ,
     \end{eqnarray}
     and 
      \begin{eqnarray} 
 \frac{h}{k_2}    &\approx & \left\{
   \begin{array}{cc}  -\log \kappa + 2 \log 2 -1 - \frac{1}{2} \kappa^2 \log \kappa + \dots  & \kappa \to 0\\
     \\
   \frac{\pi}{2}(1- \kappa)  +   \frac{\pi}{8}(1- \kappa) ^2+ \dots & \kappa\to1
      \end{array} 
       \right.\, .
          \label{eq:Asymptotic2b}
     \end{eqnarray}

\bibliographystyle{apsrev4-1}

\bibliography{superflow}
\end{document}